\documentclass[preprint,aps,amsmath,amsfonts,amssymb]{revtex4-1}

\newcommand{\be}{\begin{equation}}
\newcommand{\ee}{\end{equation}}
\newcommand{\bea}{\begin{eqnarray}}
\newcommand{\eea}{\end{eqnarray}}
\newcommand{\ba}{\begin{array}}
\newcommand{\ea}{\end{array}}

\usepackage{epsfig}
\usepackage{graphicx}
\usepackage{hyperref}

\begin{document}

\title{Generalized parton distributions in the soft-wall model of AdS/QCD }

\author{Neetika Sharma}
\affiliation{ Indian Institute of Science Education and Research Mohali,\\ S.A.S. Nagar, Mohali-140306, India.}
\begin{abstract}

We present a numerical analysis of helicity independent nucleon generalized parton distributions (GPDs) using the known formalism based on inclusion of higher Fock states in the soft-wall  approach of  the anti-de Sitter/QCD model.  We calculate the momentum space GPDs by matching the electromagnetic form factors in the AdS model to the sum rules in QCD.  We  investigate their Mellin moments, transverse impact parameter GPDs, transverse mean square radius, and transverse width.  We further extend this work to investigate the charge and anomalous magnetization densities for both unpolarized and transversely polarized nucleons. A comparison of  results on density functions  with phenomenological parametrization is also presented.
\end{abstract}
\maketitle

%11.25.Tq	Gauge/string duality
%13.40.Gp	Electromagnetic form factors
%14.20.Dh	Protons and neutrons
%12.38.Lg	Other nonperturbative calculations
%12.38.Aw	General properties of QCD
 %14.65.Bt	Light quarks
%13.60.Fz	Elastic and Compton scattering

%%%%%%%%%%%%%%%%%%%%

\section{Introduction}

Generalized parton distributions (GPDs) are fundamental quantities of theoretical and experimental endeavor in the recent past and give us essential information  about the internal structure of  nucleons  \cite{dvcs}. They represent a natural interpolation between electromagnetic form factors (EFFs) and  parton distribution functions (PDFs).  The  first moments of GPDs are related to the EFFs and  reduce to the PDFs in the forward limit. The study of these quantities is further significant to understand the issues related to the spin and orbital angular momentum of the constituents, as well as spatial structure of the nucleon.  There are several extensive reviews about GPDs in the literature \cite{dvcs,lattice}. 
GPDs are a function of the longitudinal momentum fraction of the active quark ($x$), the longitudinal momentum fraction transferred or skewness parameter ($\zeta$), and the square of the momentum transferred ($t=-Q^2$). At zero skewness ($\zeta$=0), the Fourier transform of GPDs with the momentum transfer in  transverse direction gives  impact parameter dependent GPDs \cite{impact}, which have probabilistic interpretation in terms of density functions \cite{jisum}. They provide us with information about partonic distributions in the  transverse impact parameter or  position space and give an estimate of  separation of the struck quark  and center of momentum of  the nucleon.

GPDs enter in the  measurements of amplitudes of  hard exclusive processes like deeply virtual Compton scattering and vector meson production \cite{vmd}. Various experiments, such as H1 and ZEUS at DESY \cite{desy}, COMPASS at CERN \cite{compass1},  Hall A and  CLAS at Jefferson Lab \cite{jlab1}  have measured GPDs for valence quarks.  The  upcoming experiments with high luminosity and wider kinematic range, such as, COMPASS-II \cite{compass2} and 12 GeV energy upgrade at Jefferson Lab \cite{jlab2}, will significantly advance the measurements of GPDs for sea quarks, gluons, and transverse single spin asymmetries. However, the numerical analysis is partially framework dependent and requires modeling of GPDs into functional form \cite{kelly,mrst,diehl,mstw2009}. The Euclidean lattice QCD is another important framework but the successes are limited by the uncertainties arising from the  statistical errors,  extrapolation to the physical quark mass, and complexities of numerical algorithms, etc. \cite{lattice}. Further, the dynamical observables in the Minkowski space-time are not directly obtained from the Euclidean lattice computations \cite{mlattice}. This paved the way for the formulations of alternative approaches to extract information about GPDs and  to make precise predictions in the nonperturbative regime.

Recently, light-front holography (LFH) emerged as a promising technique to unravel the structure of hadrons \cite{lfh}.  It is based on AdS/CFT correspondence between the string theory on a higher-dimensional anti$-$de Sitter (AdS) space and conformal field theory (CFT) in physical space-time to study the hadronic properties \cite{adscft}.  LFH methods were originally introduced  by matching the matrix elements of dynamical observables, e.g., electromagnetic current matrix elements in AdS space  with the corresponding expressions from light-front quantization in physical space-time \cite{adsqcd}. This provides a precise mapping of the string modes $ \Phi$ in the AdS fifth dimension to the hadron light-front wave functions $\psi$ \cite{stan2012}.  Though LFH is a semiclassical approximation for strongly coupled quantum field theories \cite{confine}, it successfully explains the general properties of mesons, e.g., mass spectra including the Regge trajectories \cite{mesonmass}, electromagnetic and gravitational form factors \cite{mesonff}, decay constants, decay widths,  distribution amplitudes \cite{mesonda}, and other physical quantities \cite{heavymeson}.  The AdS/QCD wave function has been further used to investigate the form factors, branching ratios, distribution amplitudes for the  radiative and semileptonic $B$ decays to light vector mesons \cite{ahmady}.

During the past few years significant progress has been made in the application of AdS/QCD models to baryons \cite{baryons}.  The electromagnetic form factors for the nucleon have been calculated using the nonminimal couplings \cite{abidin} and  light-front quark model with SU(6) spin flavor symmetry \cite{stansoft}.  One can also constrain the information on GPDs for valence quarks  indirectly via the sum rules that connect them with form factors. This procedure has been used to  investigate the GPDs in the  hard-wall model \cite{hard} and soft-wall model \cite{soft,dip}. It is significant to mention here that in the hard-wall model, an IR boundary $z_0=1/\Lambda_{{\rm QCD}}$ is put in the AdS space, while in the soft-wall model, a soft IR cutoff in the fifth dimension is introduced by a background  dilation  field or  confining potential \cite{models}.

Gutsche {\it et al.}  \cite{valery} have presented a variant of   the holographic soft-wall model with the inclusion of higher Fock states. The high Fock states' components  are holographically incorporated via studying the dynamics of 5D fermion fields  of different scaling dimension  in AdS space in accordance with gauge/gravity duality.  We distinguish this approach (with inclusion of higher Fock states) from the previous soft-wall model  \cite{soft} via referring to it  as the ``modified soft-wall model''.  The modified soft-wall model successfully explain the  mass spectrum and electromagnetic and axial isovector form factors for the nucleon; therefore, it  is interesting and intriguing to extend this approach to obtain GPDs using the  LFH mapping.

We perform the matching of nucleon form factors  considering the two approaches: sum rules in QCD  and  expressions obtained from the modified soft-wall model. We investigate the GPDs and their $x$-moments in the momentum space as these quantities are directly measured from the lattice QCD \cite{lattice}. The Fourier transform of GPDs to transverse position space gives the probability density for finding a valence quark at a particular transverse position inside a nucleon; this gives impetus on the calculation of GPDs in the impact parameter space. 
The transverse charge  and anomalous magnetization densities are directly connected to the Fourier transform of  EFFs, whereas the integration over parton momentum fractions ($x$)  for the GPDs yields EFFs and consequently relates the transverse charge  and anomalous magnetization density with the GPDs in impact parameter space. It is interesting to investigate the transverse charge densities for unpolarized and polarized nucleons  and measure the  spatial distribution of partons in the transverse plane.

This work is organized as follows. In Sec. \ref{formf},  we  follow the work of Ref. \cite{valery} and  outline the  essential results of EFFs in the modified soft-wall model of AdS/QCD. In Sec. \ref{gpds}, we present the  numerical results for GPDs for up and down quarks in the momentum space and also discuss their $x$-moments.  We further discuss the impact parameter dependent GPDs including their transverse mean squared radii and transverse width in Sec. \ref{impact}. Charge and  magnetization densities for both unpolarized and transversely polarized nucleons will be discussed in Sec. \ref{mag}. We also present the results for the flavor contributions of up and down quarks and a comparison with the phenomenological parametrization  in the same section. Summary and conclusions are presented in Sec. \ref{sum}.

\section{Nucleon Electromagnetic form factors and wave function} 
\label{formf}

In this section, we will reproduce the  relevant results by Gutsche {\it  et al.} \cite{valery} for the derivation of EFFs and wave functions in the modified soft-wall model. This approach is based on an action which describe hadrons with the soft-wall breaking of conformal symmetry by introducing a quadratic dilation  field $\varphi(z) =\kappa^2 z^2$. The quadratic dependence of the dilaton field  produces linear Regge-like mass spectra for hadron masses.  Also, the Dirac fermion field $\Psi(x,z)$ propagating in the 5-dimensional AdS space is with different twist dimensions which correspond to the contribution of the higher Fock state components. The nucleon structure is considered as a superposition of three valence quark states with the contribution of  higher Fock states including quarks, antiquarks, and gluons via studying the dynamics of 5-D Dirac fermion fields of different scaling dimensions in AdS space \cite{valery}.

The  AdS/QCD interaction  action which generates the nucleon form factors is expressed as
\be
S_{\rm int}^V =  \int  {\mathrm d}^4x \, {\mathrm d}z \, \sqrt{g} \, e^{-\varphi(z)} \, {\cal L}_{\rm int}^V(x,z)\,,
\ee 
where $g= |g_{MN}|$ and $\varphi(z) = \kappa^2 z^2$ is the quadratic dilaton field with $\kappa$ as the free scale parameter.  The interaction Lagrangian containing the minimal and nonminimal couplings of fermion and vector  AdS fields is given as \be
\label{lint} 
{\cal L}_{\rm int}(x,z) = \sum\limits_{i=+,-} 
\sum\limits_{\tau} \, c_\tau \, 
\bar\Psi_{i, \tau}(x,z)  \, \hat{\cal V}_i(x,z) \, \Psi_{i, \tau}(x,z)\,, 
\ee
with \bea
\hat{\cal V}_\pm(x,z)  &=& {\cal Q} \, \Gamma^M  V_M(x,z) \, \pm \, 
\frac{i}{4} \, \eta_V \,  [\Gamma^M, \Gamma^N] \, V_{MN}(x,z)  
\, \pm \, g_V\, \tau_3 \, \Gamma^M \, i\Gamma^z \, V_M(x,z)  \,.
\eea
Here  $\Psi_{\pm, \tau}(x,z)$ is the five Dimensional fermion fields with spin $J=1/2$ and scaling dimension $\tau$; $V_M(x,z)$  is the vector fields which is holographic dual of  the electromagnetic field; $V_{MN} = \partial_M V_N - \partial_N V_M$ is the stress tensor of the vector field; ${\cal Q} = {\rm diag}(1, 0)$ is the nucleon charge matrix, $\tau_3 = {\rm diag}(1, -1)$ is the Pauli isospin matrix;  $\Gamma^M = \epsilon_a^M \Gamma^a$ and  
$\Gamma^a = (\gamma^\mu, - i\gamma^5)$ are the five-dimensional Dirac matrices.

Following Ref. \cite{valery}, the expressions for Dirac and Pauli nucleon form factors  are given as 
\bea
F_1^p(Q^2) &=& C_1(Q^2) + g_V C_2(Q^2)
+ \eta_V^p C_3(Q^2)\,,\nonumber\\
F_2^p(Q^2) &=& \eta_V^p C_4(Q^2)\,,\nonumber\\
F_1^n(Q^2) &=& - g_V C_2(Q^2) + \eta_V^n C_3(Q^2)\,,
\nonumber\\
F_2^n(Q^2) &=& \eta_V^n C_4(Q^2), 
\label{fie}
\eea
where $C_i(Q^2)$ are the structure integrals expressed as
\bea 
C_{1}(Q^2) &=& \frac{1}{2} \, \int\limits {\mathrm d}z \, V(Q^2,z) 
\ \sum\limits_\tau \, c_\tau \, 
\biggl( [f^L_{\tau}(z)]^2 + [f^R_{\tau}(z)]^2 \biggr)\,, 
\nonumber\\ 
C_{2}(Q^2) &=& \frac{1}{2} \, \int\limits  {\mathrm d}z \, 
V(Q^2, z) \ \sum\limits_\tau \, c_\tau \, 
\biggl( [f^R_{\tau}(z)]^2 - [f^L_{\tau}(z)]^2 \biggr)\,, 
\nonumber\\  
C_{3}(Q^2) &=& \frac{1}{2} \, \int\limits  {\mathrm d} z z \, \partial_z V(Q^2, z) 
\ \sum\limits_\tau \, c_\tau \, 
\biggl( [f^L_{\tau}(z)]^2 - [f^R_{\tau}(z)]^2 \biggr)\,, 
\nonumber\\  
C_{4}(Q^2) &=& 2 m_N \, \int\limits  {\mathrm d}z z \, V(Q^2,z) 
\ \sum\limits_\tau \, c_\tau \, 
f^L_{\tau}(z) f^R_{\tau}(z)\,. \label{Ci} 
\eea
The functions  $f^{L}_{\tau}(z)$ and $f^{R}_{\tau}(z)$ are the bulk profiles of fermions corresponding to the left-and right- handed ground-state nucleons  with radial quantum number $n=0$.  
The ground-state  nucleon wave functions are expressed as
\bea
f^L_{\tau}(z) &=& \sqrt{\frac{2}{\Gamma(\tau)}} 
\ \kappa^{\tau}
\ z^{\tau-1/2} \ e^{-\kappa^2 z^2/2} \,, \label{flr1} \\
f^R_{\tau}(z) &=& \sqrt{\frac{2}{\Gamma(\tau-1)}} 
\ \kappa^{\tau-1} \ z^{\tau-3/2} \ e^{-\kappa^2 z^2/2} \,. 
\label{flr2}
\eea

The $V(Q^2,z)$ is the bulk-to-boundary propagator of the transverse massless vector bulk field in terms of the gamma function $\Gamma(a)$ and Tricomi confluent hypergeometric function  $U(a,b,z)$ \cite{bulk}
\be
 V(Q^2,z)=\Gamma \left(1+ {Q^2 \over 4 \kappa^2 }\right) 
 U\bigg({Q^2 \over 4 \kappa^2},\,0,\,\kappa^2 z^2 \bigg)\,.
 \label{prop}
 \ee
The propagator in Eq. (\ref{prop})  can be conveniently written in an  integral representation \cite{bulk}  \be V(Q^2,z)=\kappa^2\, z^2\int\limits_0^1 \frac{ {\mathrm d} x}{(1-x)^2} x^{Q^2 \over 4 \kappa^2} e^{-\kappa^2 z^2 x/(1-x)}\,.  \label{btb} \ee The bulk-to-boundary propagator $V(Q^2,z)$ satisfies the normalization condition and follows the ultraviolet boundary $V(Q^2,0)=1$ and infrared boundary $V(Q^2, \infty)=0$ condition. The analytical expressions for nucleon form factors are obtained in Ref. \cite{valery} after substituting the hadronic states with twist $\tau$ dimensions  Eqs. (\ref{flr1})-(\ref{flr2}), the integral representation of the bulk-to-boundary propagator Eq. (\ref{btb}),  in the structure integrals Eq. (\ref{Ci}). It has already been proven to give  satisfactory agreement to data on nucleon form factors with the minimum number of free parameters \cite{valery}.

%%%%%%%%%
\section{Generalized Parton Distributions and their moments}
\label{gpds}

In this section, we calculate the GPDs for the nucleon using the correspondence procedure based on light-front holography. We perform a matching of the matrix elements for nucleon form factors considering two approaches: one is  sum rules in QCD  and the other is expressions obtained in the modified soft-wall model of AdS/QCD with arbitrary twist dimensions.  The sum rules relate the GPDs for unpolarized quarks with the form factors \cite{jisum}
\bea
 \label{sumrule1}
F_1^q(Q^2) &=& \int_{0}^1 dx\; H_v^q(x,  Q^2) \,,
\\
F_2^q(Q^2) &=& \int_{0}^1 dx\; E_v^q(x, Q^2) \,. 
\label{sumrule2}
\eea
We have defined the GPDs for valence quarks (minus antiquark) as  $H_v^q(x, Q^2)=H^q(x,0,Q^2) +H^q(-x,0,Q^2)$ and $E_v^q(x,Q^2)=E^q(x,0,Q^2)+E^q(-x,0,Q^2)$ for the zero skewness. The Dirac and Pauli form factors for the nucleon are  given by charge weighted sum  \be F_{i}^{N}(Q^2) = \sum_{q} e_q^{N}\, F^q_{i}(Q^2) \,, \label{sumrule} \ee with appropriate coefficients $e^p_u =e^n_d = {2\over 3}$,  and $e^p_d=e^n_u= -{1 \over 3}$.   We restrict ourselves to the contribution from the valence  quarks only, whereas the contributions of heavier strange and charm quarks have been ignored.

\begin{figure}[h]
\begin{minipage}[c]{0.98\textwidth}
\small{(a)}
\includegraphics[width=7cm,height=5cm,clip]{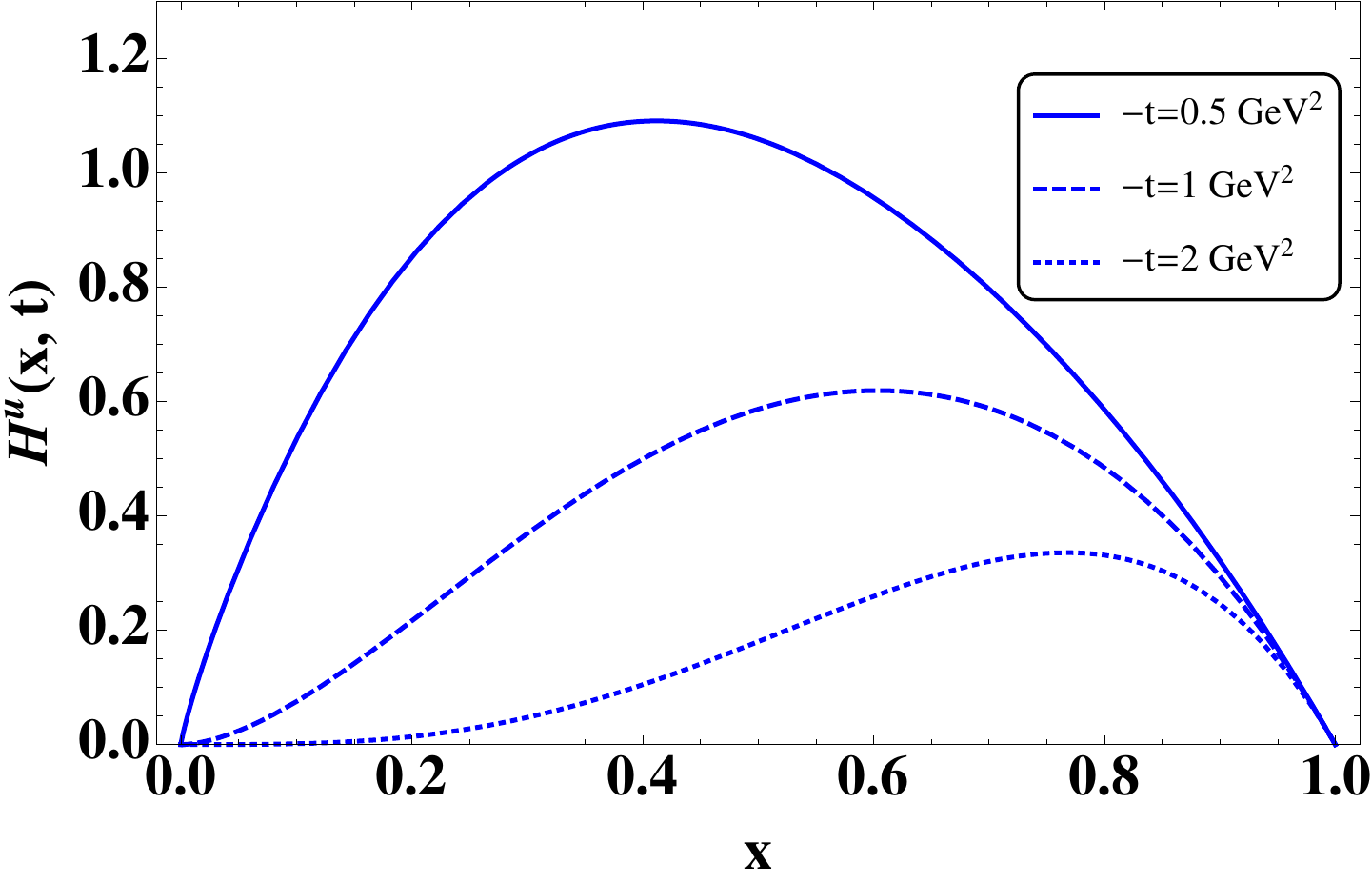}
\small{(b)}
\includegraphics[width=7cm,height=5cm,clip]{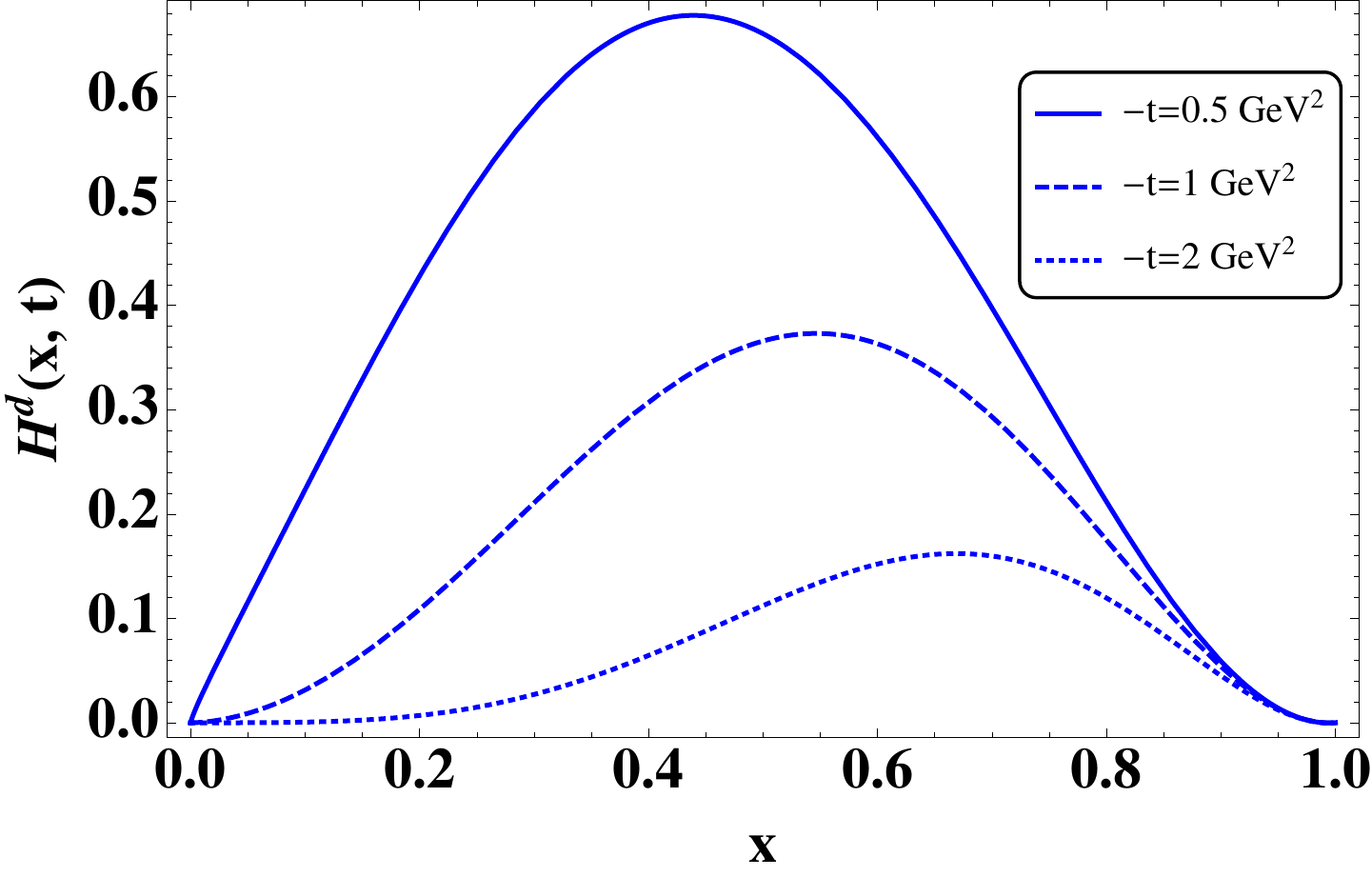}
\end{minipage}
\begin{minipage}[c]{0.98\textwidth}
\small{(c)}\includegraphics[width=7cm,height=5cm,clip]{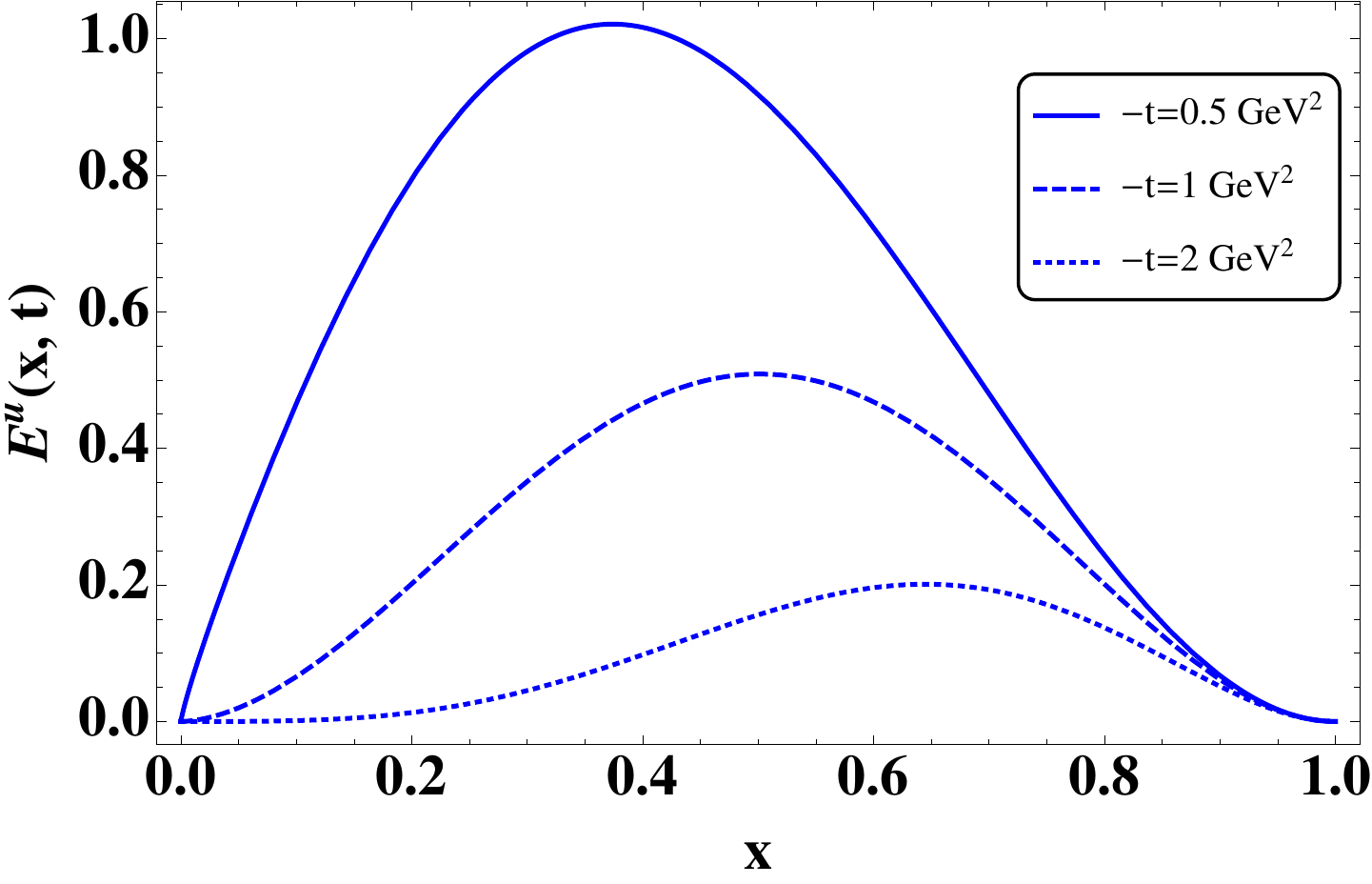}
\small{(d)}\includegraphics[width=7cm,height=5cm,clip]{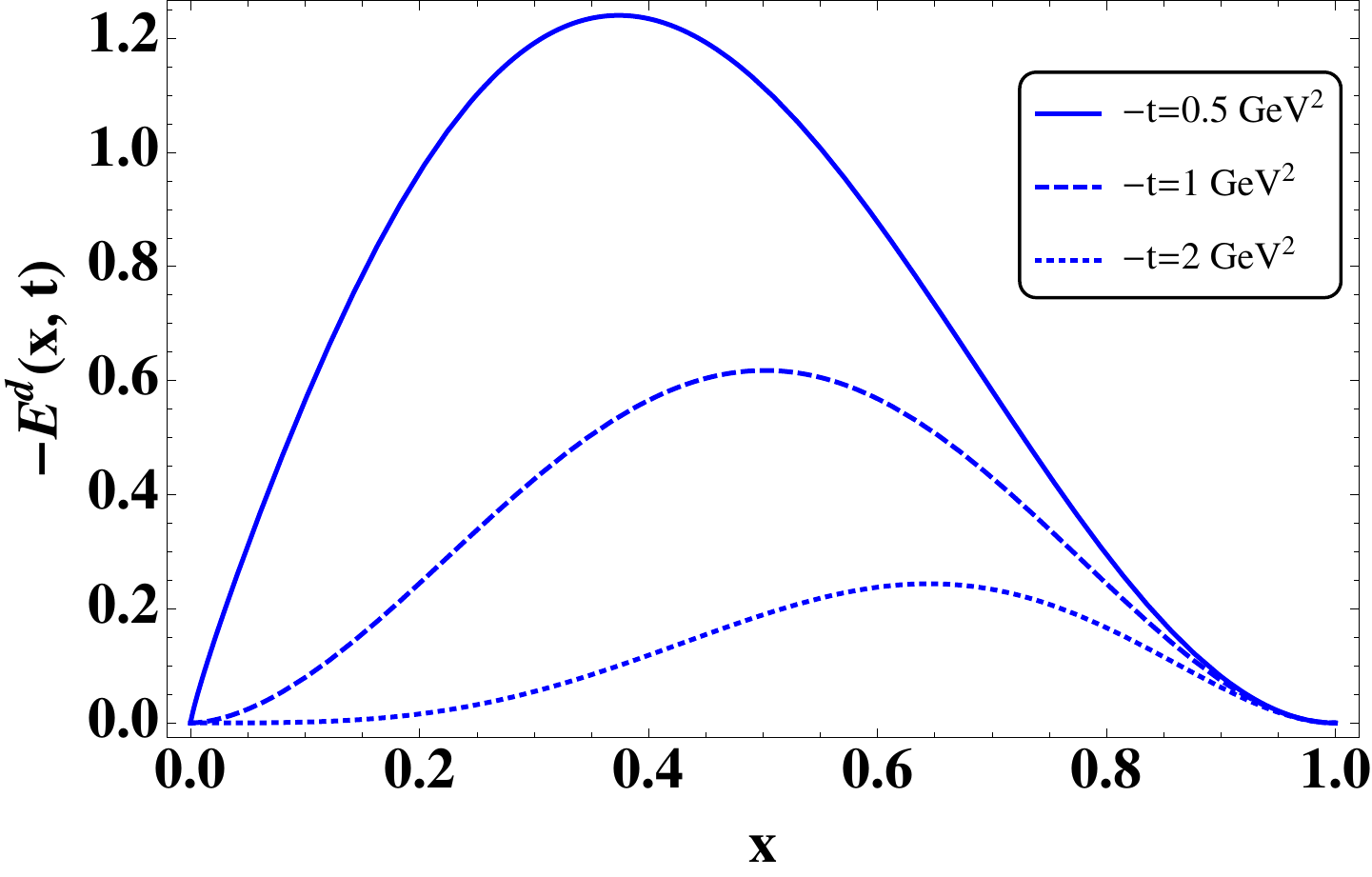}
\end{minipage}
\caption{\label{hu}Plots of (a) the generalized parton distributions $H^u(x,t)$ vs $x$ for fixed values of $ -t = Q^2$ for up quark, (b) $H^d(x,t)$ vs $x$ for fixed values of $-t$ for down quark, (c) $E^u(x,t)$ vs $x$ for fixed values of $ -t$ for up quark, and (d) $E^d(x,t)$ vs $x$ for fixed values of $-t$ for down quark.}
\end{figure}

The  explicit expressions for up and down quark  GPDs in the modified soft-wall model  by exploiting the integral representation of bulk-to-boundary propagator $V(Q^2,z)$ are given as
\bea
H^q (x,Q^2) &=&  \sum\limits_\tau \, c_\tau \, q(x, \tau)\, x^{Q^2/ 4 \kappa^2} \,,  \nonumber\\
E^q (x, Q^2) &=&  \sum\limits_\tau \, c_\tau \,  {e^q}(x, \tau) \, x^{Q^2/ 4 \kappa^2} \,.
\eea
The quark distribution functions  $q(x, \tau)$ and ${e^q}(x, \tau)$ are
\bea 
q(x, \tau) &=& \alpha_1^q \gamma_1(x, \tau) + \alpha_2^q\gamma_2(x, \tau)+\alpha_3^q \gamma_3(x, \tau) \,, \nonumber\\
{e^q}(x, \tau) &=& \alpha_3^q \gamma_4(x, \tau) \,.
 \eea
The flavor coupling parameters $\alpha^q_i$ and $\gamma_i(x)$ are expressed as
\bea \alpha_1^u =2, ~ \alpha_2^u= g_V, ~\alpha^u_3 = 2 \eta^p_V + \eta^n_V \,, \nonumber \\
 \alpha_1^d =1, ~ \alpha_2^d =-g_V, ~\alpha^d_3 = \eta^p_V +2 \eta^n_V\,, \eea
and \bea 
\gamma_1(x, \tau) &=& -\frac{1}{2}  \,   { (1 - 2 \tau + x \tau )  (1-x)^{\tau -2} } \,,\nonumber\\
\gamma_2(x, \tau) &=& \frac{1}{2}  \,  { (1 - x  \tau )  (1-x)^{\tau -2} } \,,\nonumber\\
\gamma_3(x, \tau) &=&  {  (1 -3 x \tau + x^2 \tau + x^2 \tau^2  )  (1-x)^{\tau -2} }\,, \nonumber \\
\gamma_4(x, \tau)&=& {2 M_n \over \kappa} \, {\, \tau \, \sqrt{\tau-1}  (1-x)^{\tau -1} }
\,.
\eea
We consider the three leading order dimensions (twist $\tau =3, 4, 5$) which correspond to the contribution of quarks, antiquarks, and gluons. It is important to mention here that  for twist  $\tau =3$, these results are the same as predictions of the soft-wall model \cite{soft}.  In Figs. \ref{hu}(a)$-$\ref{hu}(b), we have presented the behavior of spin conserving GPD $H^{u/d}(x,t=-Q^2)$ as a function of $x$ for different values of $t = -0.5, -1,-2$ GeV$^2$  for up  and down quarks. The qualitative behavior of  GPD  is the same for both quarks. The profile function increases with $x$, obtains a maxima and then falls to zero as $x \to 1$. It is interesting to note that the falloff behavior is faster for the down quark. 
In Figs. \ref{hu}(c)$-$\ref{hu}(d), we present the spin changing GPDs $E^{u/d}(x,t)$ as a function of $x$ for different values of $t$ for the up and down quarks. In this case also the GPDs  increase to a maximum value and then decrease, however, the falloff behavior with $x$  is the same for both up and down quarks. For all cases the peak of GPDs shifts towards a higher value of $x$ for the larger value of momentum transferred $t$  as the struck parton with higher momentum is more likely to have a higher value of $x$.

%%%%%%%%%%%%%%%%%%%%%%%%%%%%%%%%%%%%%%%%%%%%%% Plots Moments of GPDs
\begin{figure}[htbp]
\begin{minipage}[c]{0.98\textwidth}
\small{(a)}
\includegraphics[width=7cm,height=5cm,clip]{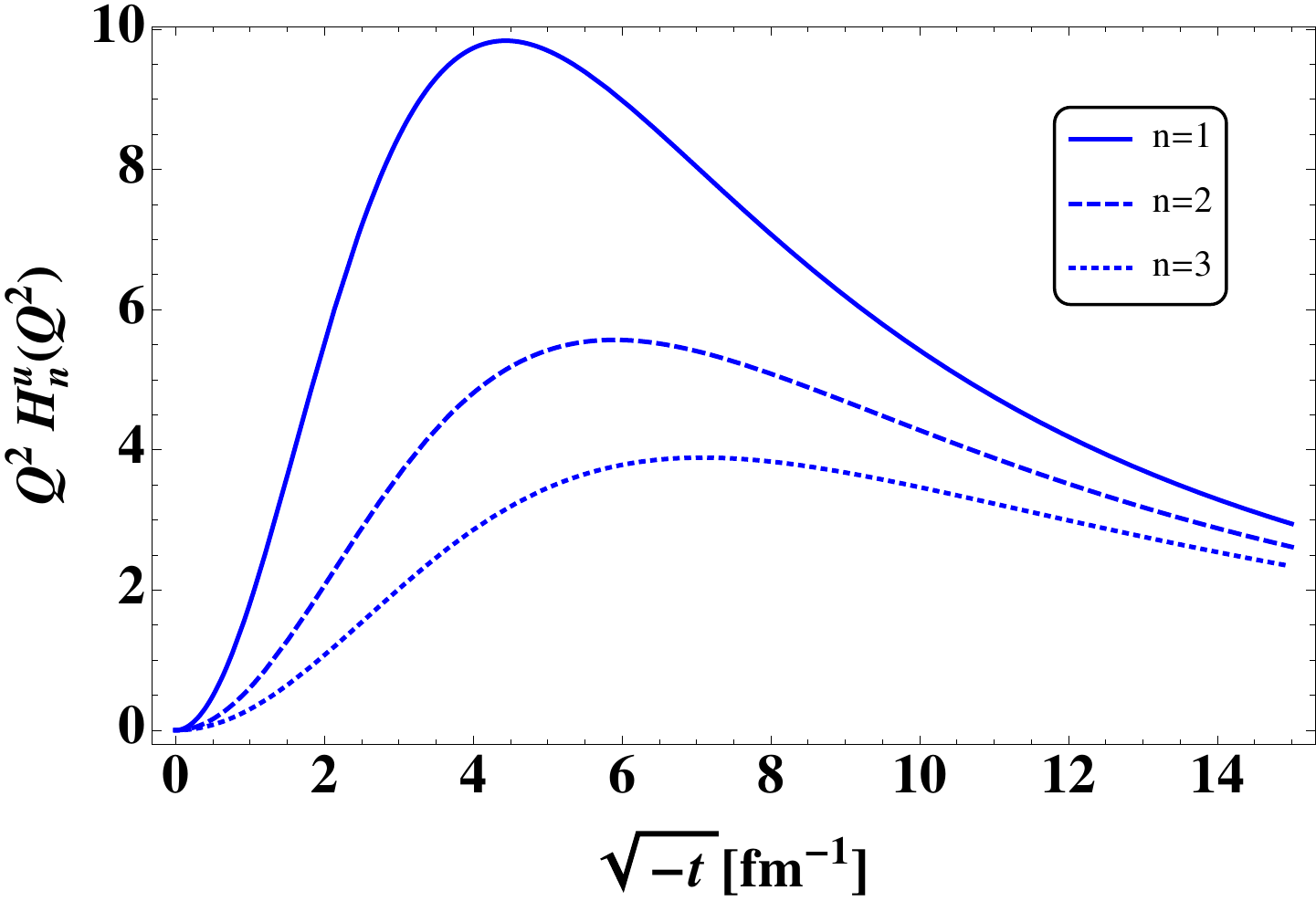}
\small{(b)}
\includegraphics[width=7cm,height=5cm,clip]{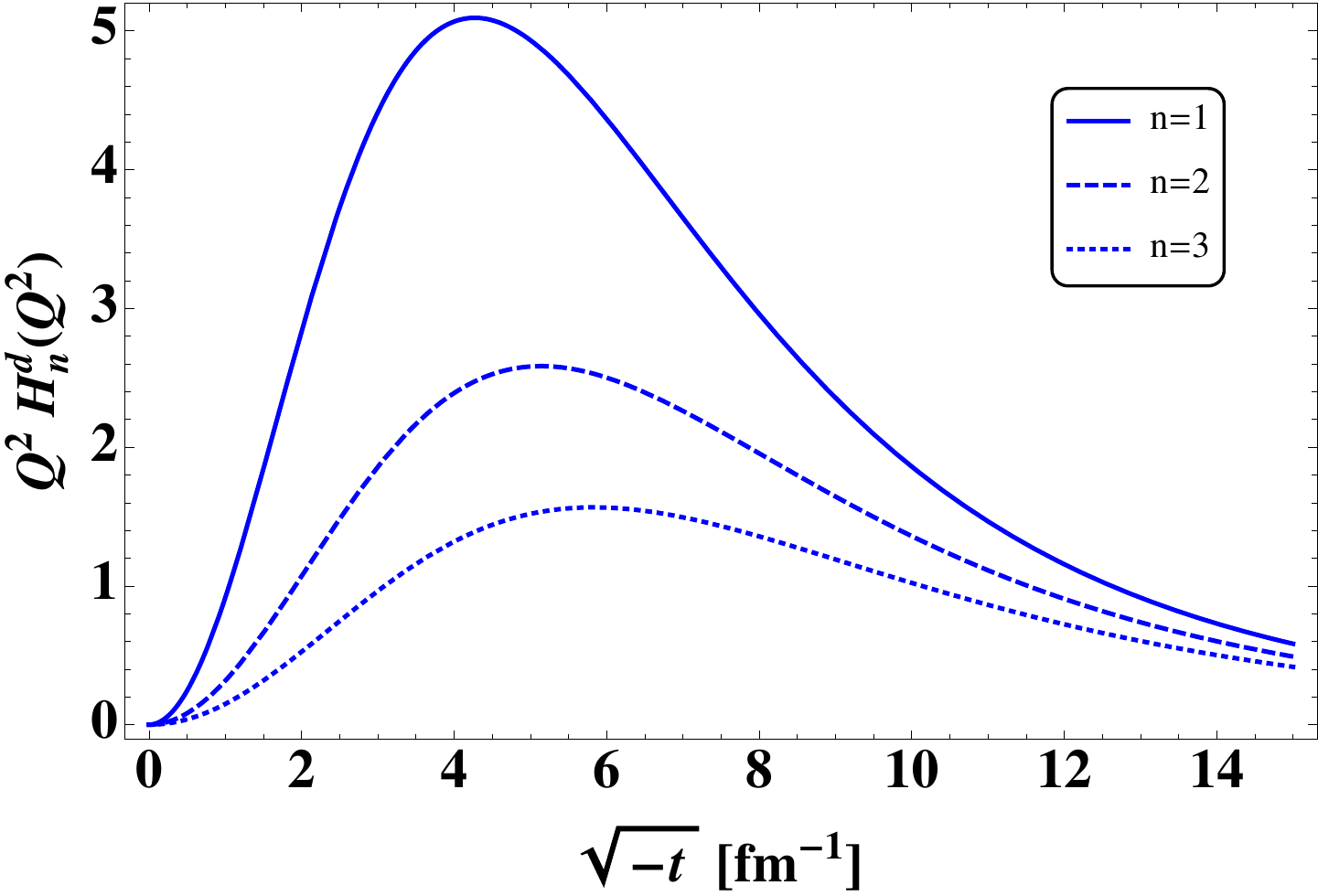}
\end{minipage}
\begin{minipage}[c]{0.98\textwidth}
\small{(c)}\includegraphics[width=7cm,height=5cm,clip]{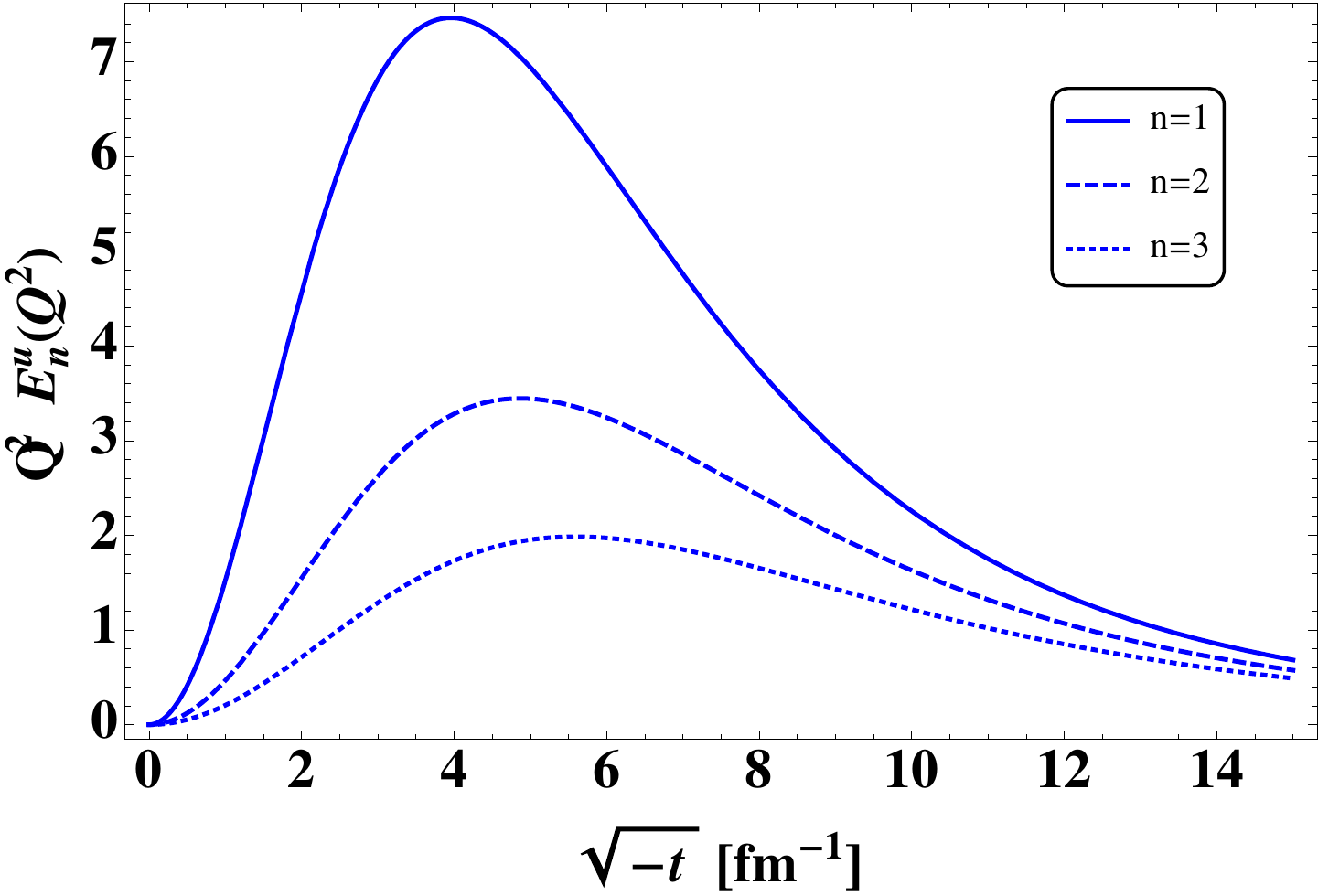}
\small{(d)}\includegraphics[width=7cm,height=5cm,clip]{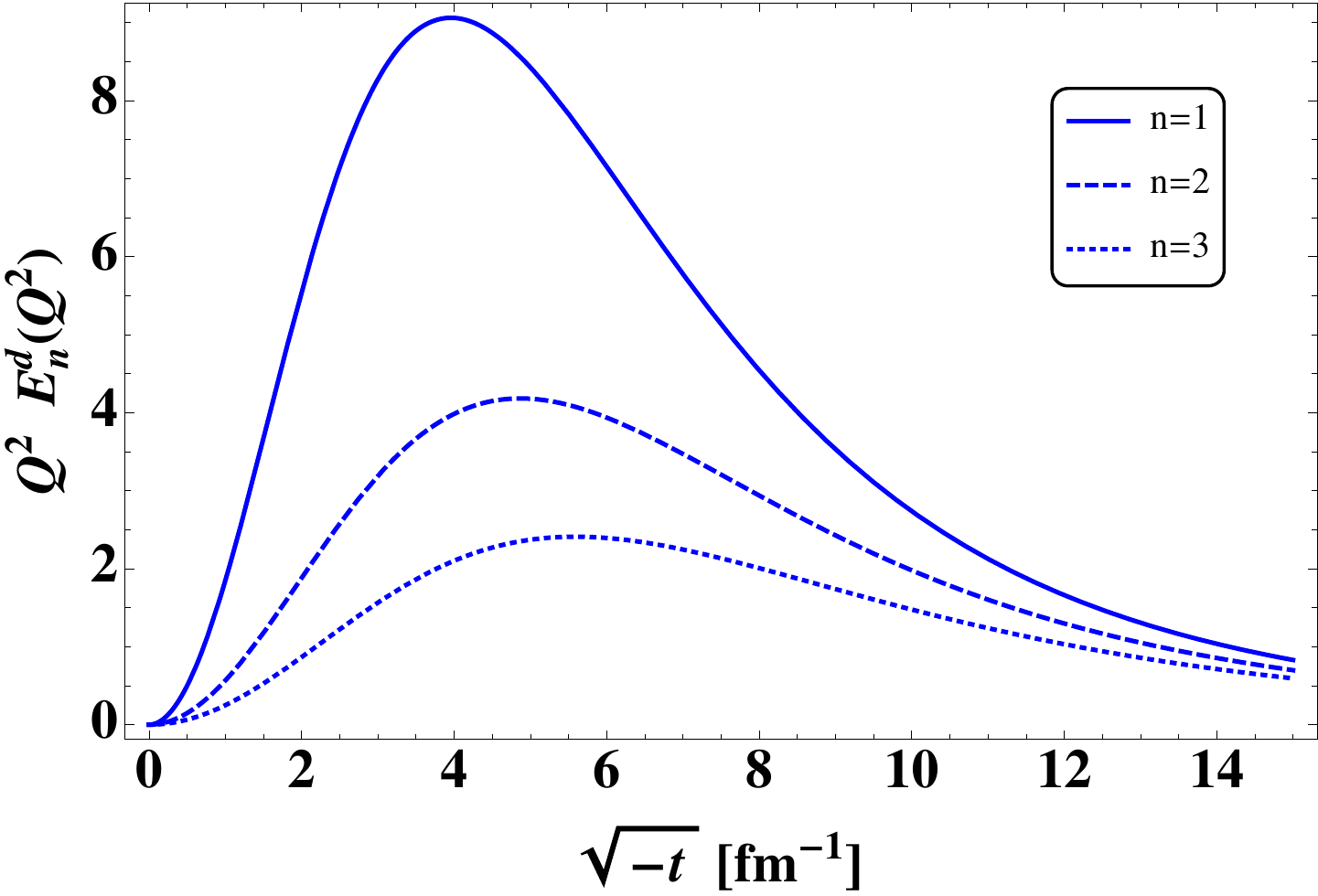}
\end{minipage}
\caption{\label{mhu} Plots of first three moments of (a) generalized parton distribution $H_n^u(x,t)$ vs ${\sqrt -t}$  for up quark; (b) $H_n^d(x,t)$ vs  ${\sqrt -t}$  for down quark; (c) $ GPD E_n^u(x,t)$ vs  ${\sqrt -t}$ for up quark; and (d) $E_n^d(x,t)$ vs  ${\sqrt -t}$  for down quark.}
\end{figure}

We now use these GPDs to compute higher order moments in $x$ for the valence GPDs defined as $H^q_n (Q^2)$ and $E^q_n (Q^2)$:
\bea 
H^q_n (Q^2) &=&  \int {\mathrm d} x\, x^{n-1} H^q (x,Q^2) \,, \nonumber \\
E^q_n (Q^2) &=&  \int {\mathrm d} x\, x^{n-1} E^q (x,Q^2) \,.
\eea
Integrating over the parameter $x$ give the moments of GPDs:
\bea 
H^q_n (Q^2) &=&  \alpha_1^q  \sum \limits_\tau \, c_\tau   \beta^1_n (Q^2, \tau) + 
\alpha_2^q \sum \limits_\tau \, c_\tau  \beta^2_n(Q^2, \tau)+
\alpha_3^q \sum \limits_\tau \, c_\tau  \beta^3_n(Q^2, \tau)  \,, \nonumber \\
E^q_n (Q^2) &=& \alpha_3^q \, \sum \limits_\tau \, c_\tau \,  \beta^4_n(Q^2, \tau) \,.
 \eea 
Here  we have defined the parameters $\beta_i(Q^2, \tau)$ in terms of the beta functions $B(m,n)$ %= {\Gamma(m) \Gamma (n) \over \Gamma(m+n)}$ 
\bea 
\beta^1_n(Q^2, \tau) &=& { 1\over 2} \left( 2 \tau + {a+n-1 } \right)  B(a+n, \tau)
\,,\nonumber\\
\beta^2_n(Q^2, \tau) &=&  - { 1\over 2} {\left( a+n-1 \right) }  B(a+n, \tau)
\,,\nonumber\\
\beta^3_n(Q^2, \tau) &=& ( a+n-1)  \left( { a+n-1}-{ a +n \over \tau} \right)  B(a+n, \tau+1)
\,, \nonumber \\
\beta^4_n(Q^2, \tau) &=&  {2 M_n \over \kappa} \tau  {\sqrt {\tau-1}}\,  {B} (a+n, \tau)
\,.
\eea
 The first moments of GPDs give the EFFs discussed in Eqs. (\ref{sumrule1})-(\ref{sumrule2}), the second moments $H^q_2$ and $E^q_2$ correspond to gravitational form factors, and the third moments $H^q_3$ and $E^q_3$ give form factors of a twist-two operator containing two covariant derivatives. The higher order moments generate the form factors of higher-twist operators.
In Figs. \ref{mhu}(a)$-$ \ref{mhu}(b),  we have plotted the behavior of the first  three moments of GPD  $Q^2 H_n^{u/d}(Q^2)$ with  momentum $\sqrt{-t}$  for up and down quarks. In Figs. \ref{mhu}(c)$-$\ref{mhu}(d), we have shown the behavior of first  three moments of GPDs  $Q^2 E_n^{u/d}(Q^2)$  with $\sqrt{-t}$ for  up and down quarks. We observe that the qualitative behavior of moments of GPD is same for  up and down quarks. 
The overall behavior of GPD moments with $t$ is the same as the behavior of profile functions with momentum fraction $x$. We also observed that the variation of the moments with $t$ becomes slower as index $n$ increases.  This can be understood in terms of  a decrease of the profile functions with momentum fraction $x$, which results in a weaker  $t$ slope for higher values of $x$. A similar trend has been observed in lattice QCD calculations of GPD moments \cite{mlattice}.

%%%%%%%%%%%%%%%%%%%%%%%%%%%%%%%%%%%%%%%%%%%%%%%%%
\section{ GPDs in impact parameter space}
\label{impact}
GPDs in the momentum space are related to their impact parameter dependent parton distribution by the Fourier transform \cite{impact}. The impact parameter GPDs  give the probability density for finding a quark with a longitudinal momentum fraction ($x$) and transverse position ($b_{\perp}$) in a nucleon, minus the corresponding probability density for antiquarks for both the parton and nucleon being unpolarized. By definition, the GPDs in transverse impact parameter space are given as \cite{impact, diehl}
\bea
q(x, b_{\perp})  &=& \int { {\mathrm d}^2 
q_{\perp} \over {(2 \pi) }^2  } \, e^{\iota b_{\perp}. q_{\perp } } H(x,  q^2)\,, \label{imp1} \\
e^q (x, b_{\perp})  &=& \int { {\mathrm d}^2 
q_{\perp} \over {(2 \pi) }^2  } \, e^{\iota b_{\perp}. q_{\perp } } E(x,  q^2)\,.
\label{imp2}\eea
In the modified soft-wall model,  the expressions for GPDs in transverse impact parameter space are
\bea
q(x, b_{\perp})  &=&  \sum\limits_\tau \, c_\tau \, q(x, \tau)\, {\kappa^2 \over \pi \log(1/x)}  e^{b_\perp^2 \kappa^2 \over \log(x) } \,, \nonumber \\
e^q(x, b_{\perp})  &=&  \sum\limits_\tau \, c_\tau \, e^q(x, \tau)\, {\kappa^2 \over \pi \log(1/x)}  e^{b_\perp^2 \kappa^2 \over \log(x) } \,. \eea

\begin{figure}[htbp]
\begin{minipage}[c]{0.98\textwidth}
\small{(a)}
\includegraphics[width=7cm,height=5cm,clip]{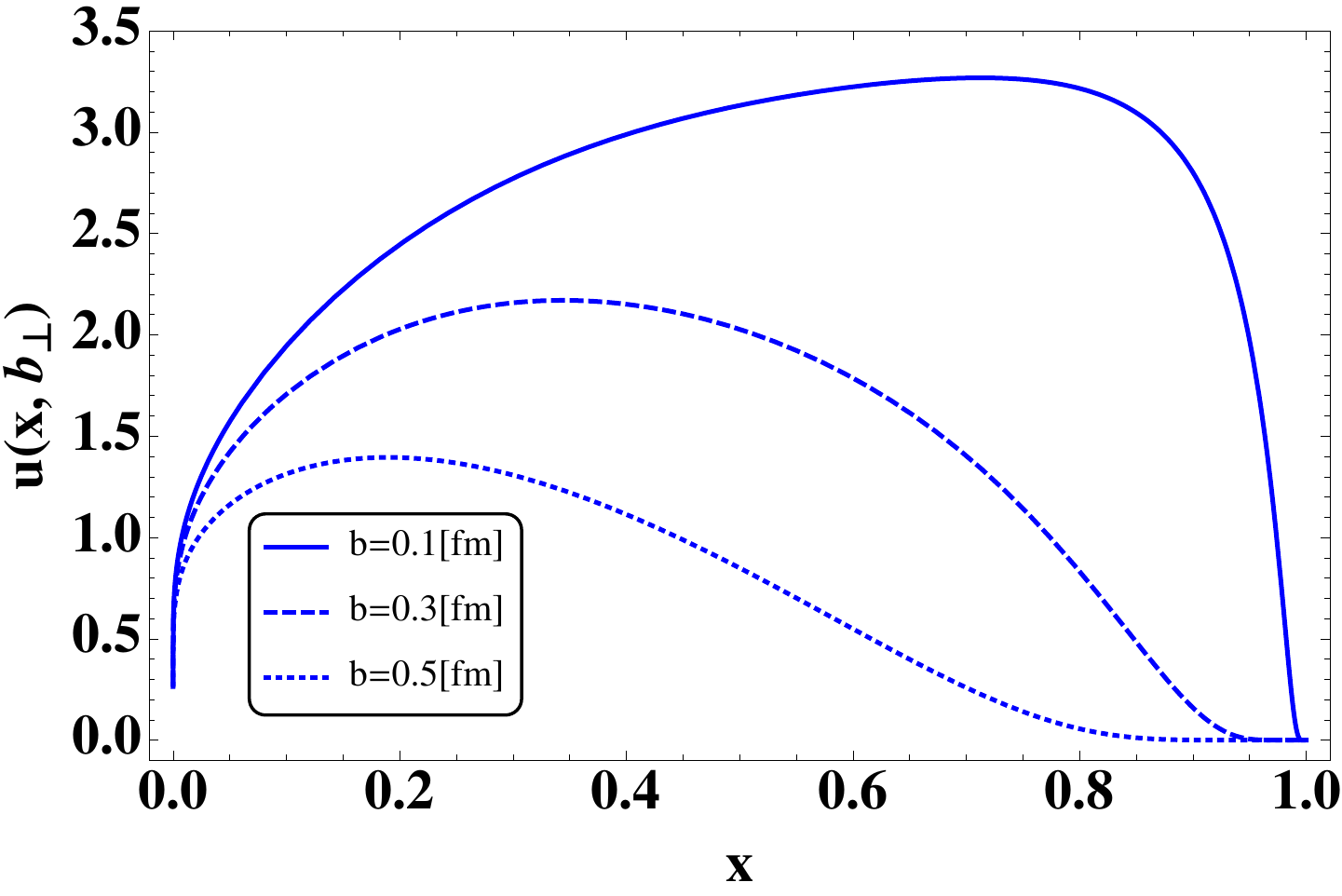}
\hspace{0.1cm}
\small{(b)}\includegraphics[width=7cm,height=5cm,clip]{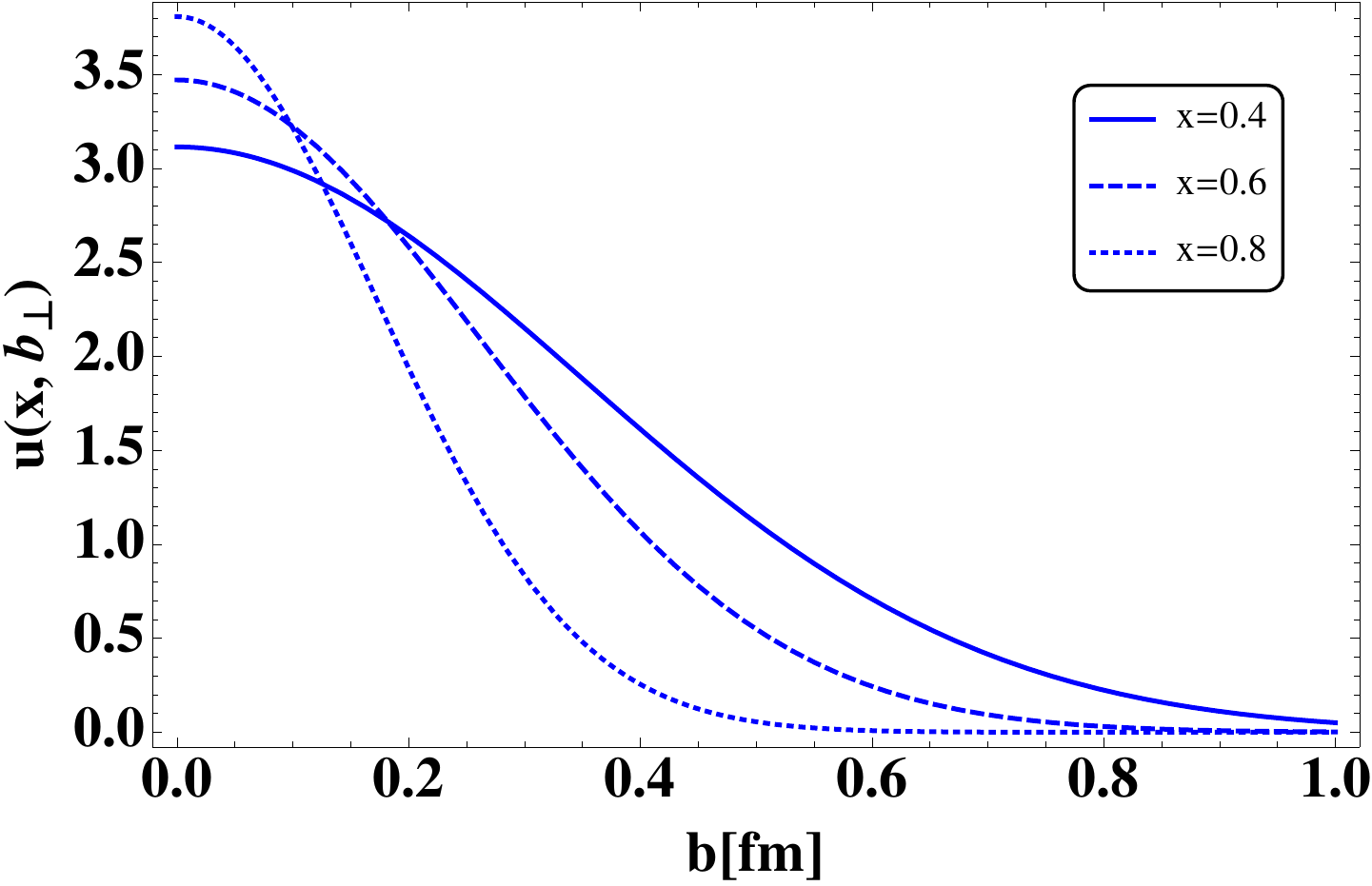}
\end{minipage}
\begin{minipage}[c]{0.98\textwidth}
\small{(c)}\includegraphics[width=7cm,height=5cm,clip]{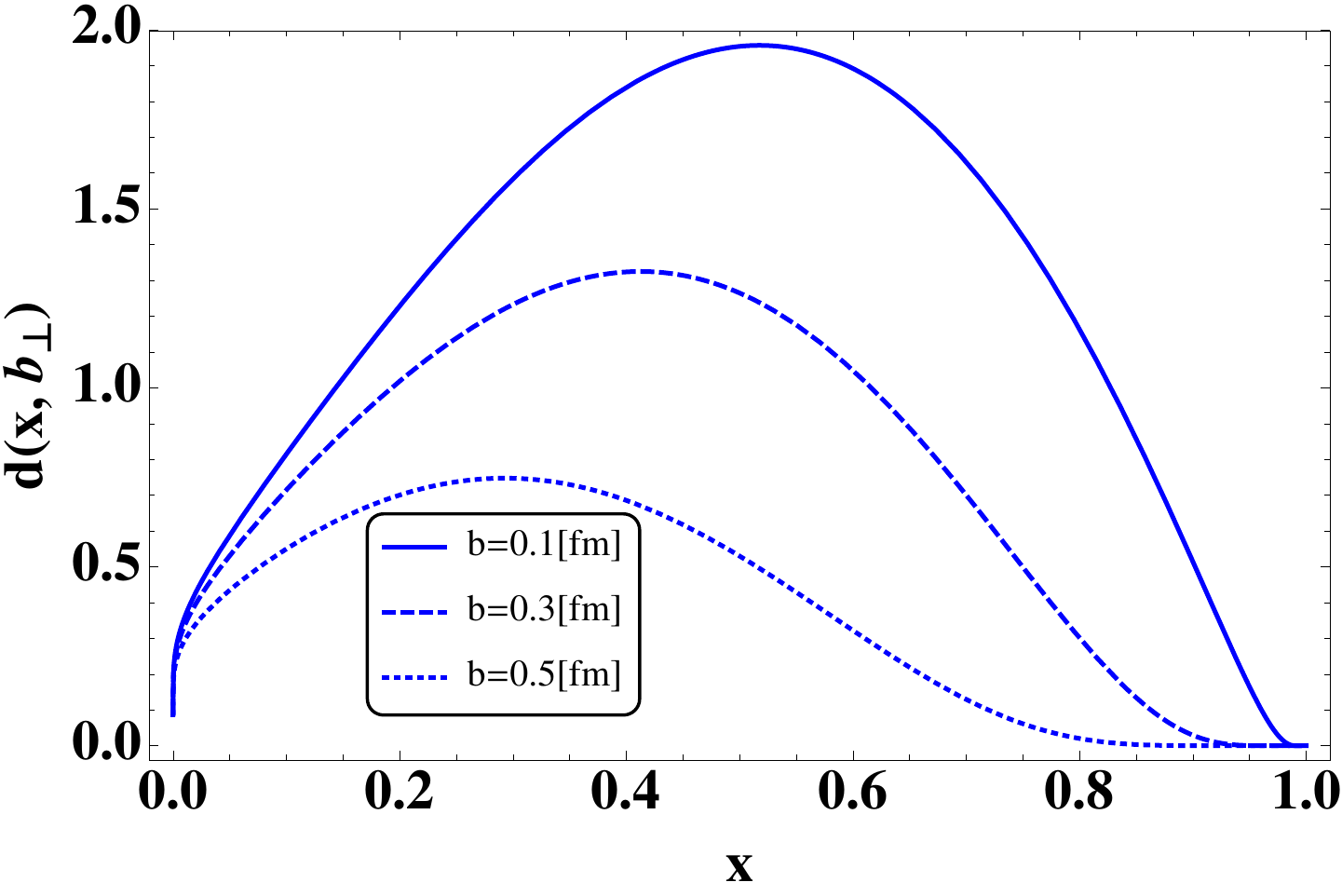}
\hspace{0.1cm}
\small{(d)}\includegraphics[width=7cm,height=5cm,clip]{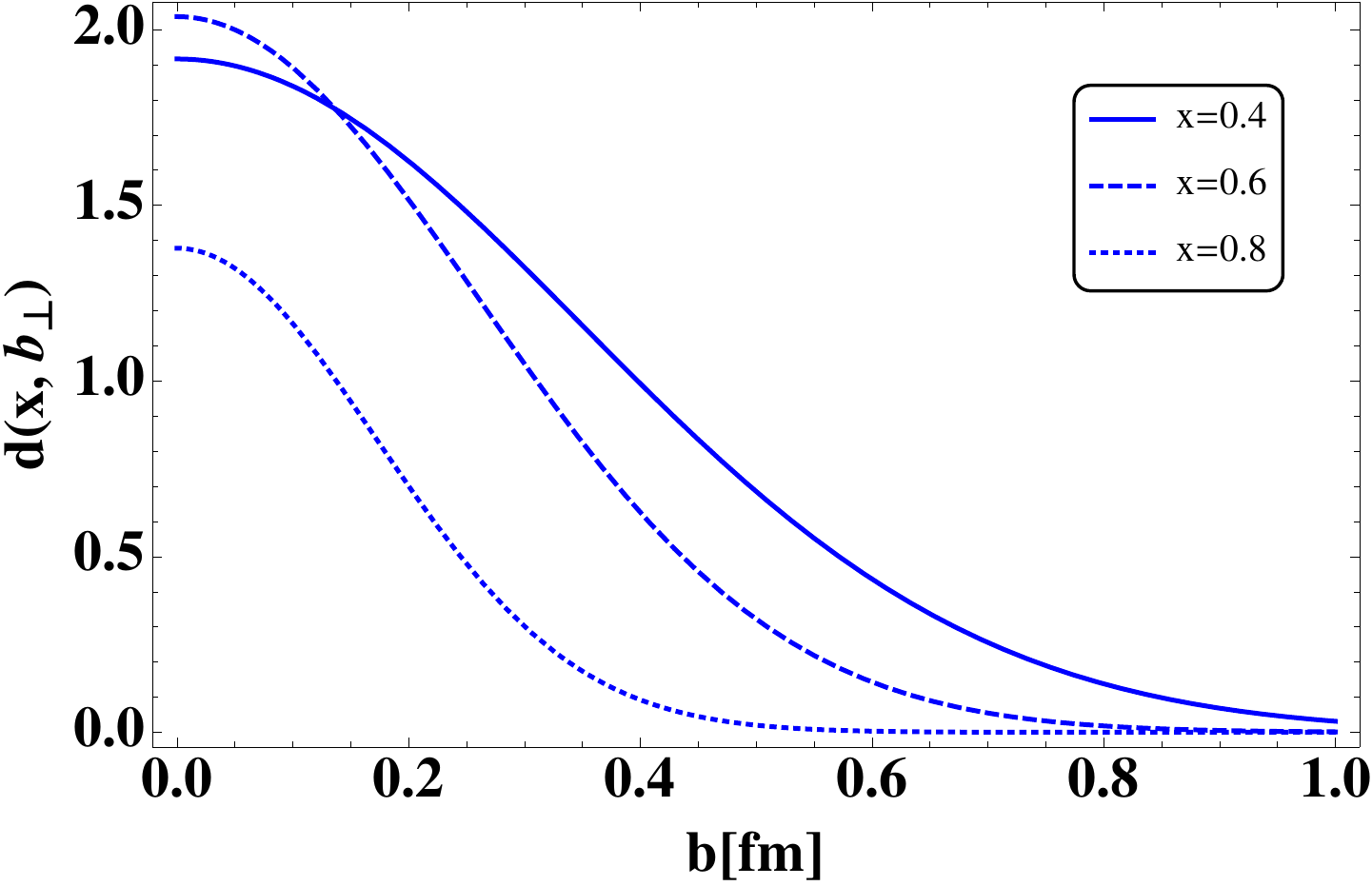}
\end{minipage}
\caption{\label{hub}(Color online) Plots of (a)  $u(x,b_\perp)$ vs $x$ for fixed values of $b=0.1, 0.3, 0.5$ fm for up quark;  (b)  $u(x,b_\perp)$ vs the impact parameter $b = |b_\perp|$ for fixed value  of $x=0.4,0.6,0.8$ for up quark; (c) $d(x,b_\perp)$ vs $x$ for fixed values of $b$  for down quark; and (d) $d(x,b_\perp)$ vs  $b$ for fixed value  of $x$ for down quark.}
\end{figure}

\begin{figure}[htbp]
\begin{minipage}[c]{0.98\textwidth}
\small{(a)}
\includegraphics[width=7cm,height=5cm,clip]{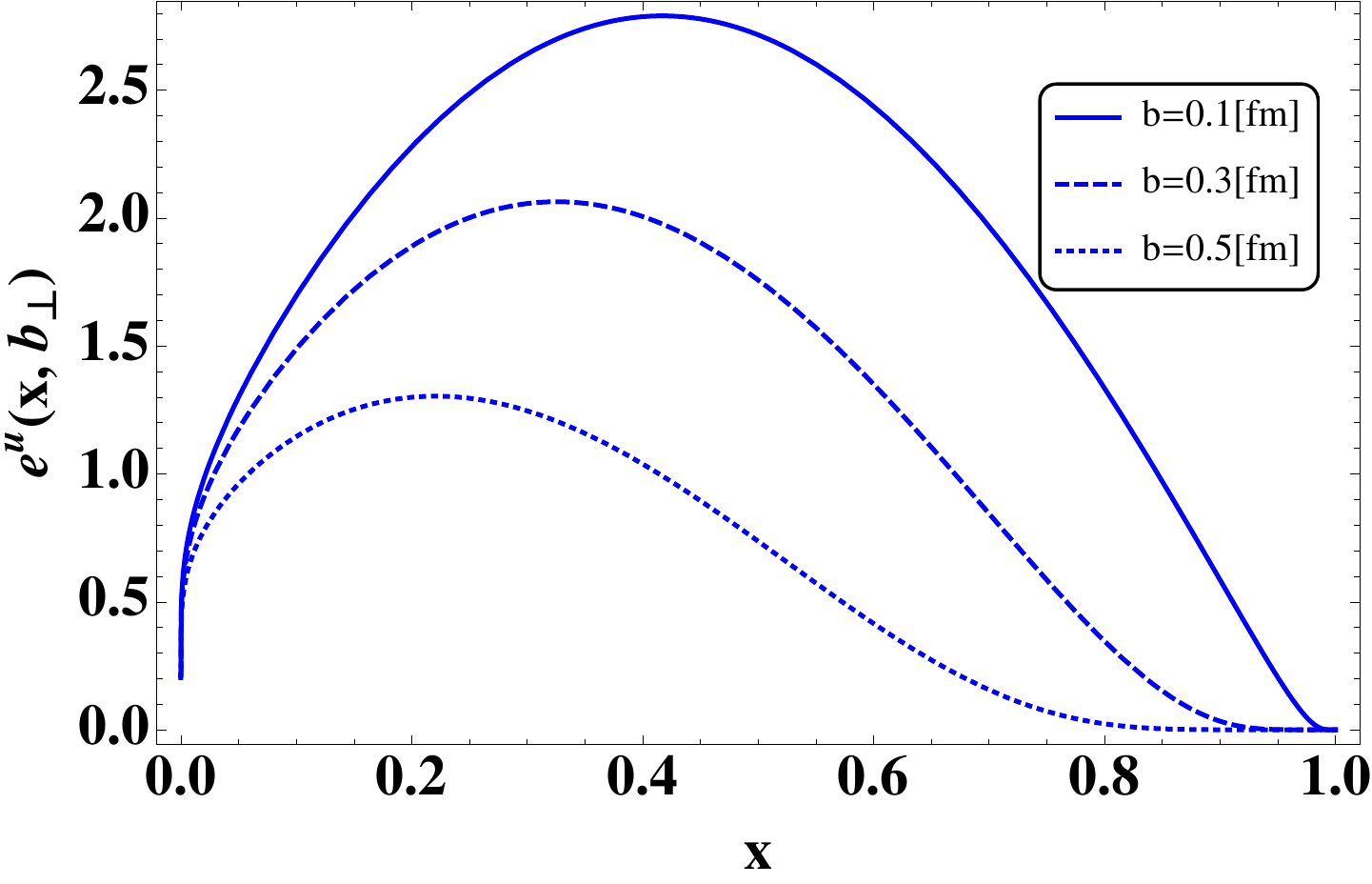}
\hspace{0.1cm}
\small{(b)}\includegraphics[width=7cm,height=5cm,clip]{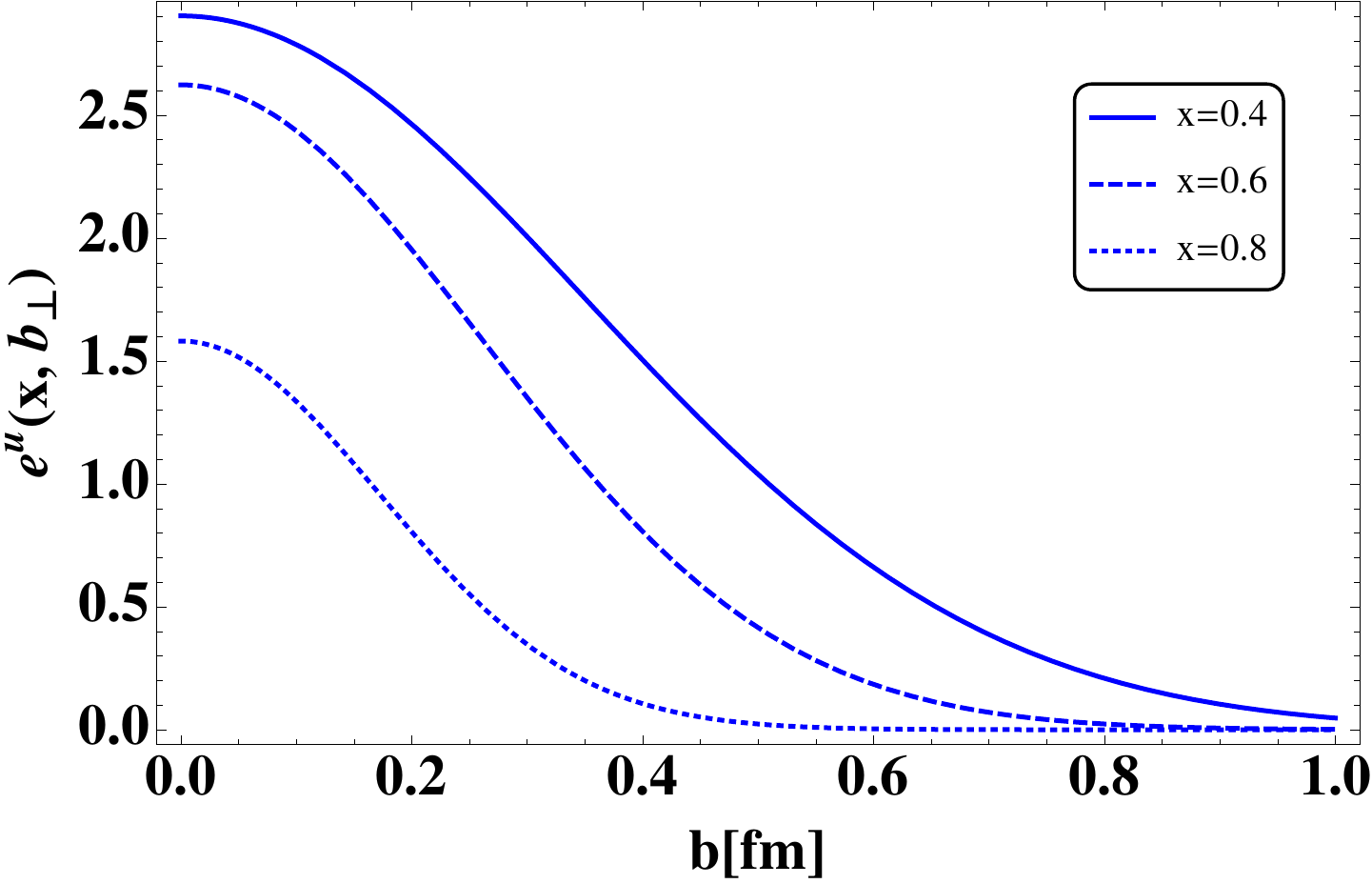}
\end{minipage}
\begin{minipage}[c]{0.98\textwidth}
\small{(c)}\includegraphics[width=7cm,height=5cm,clip]{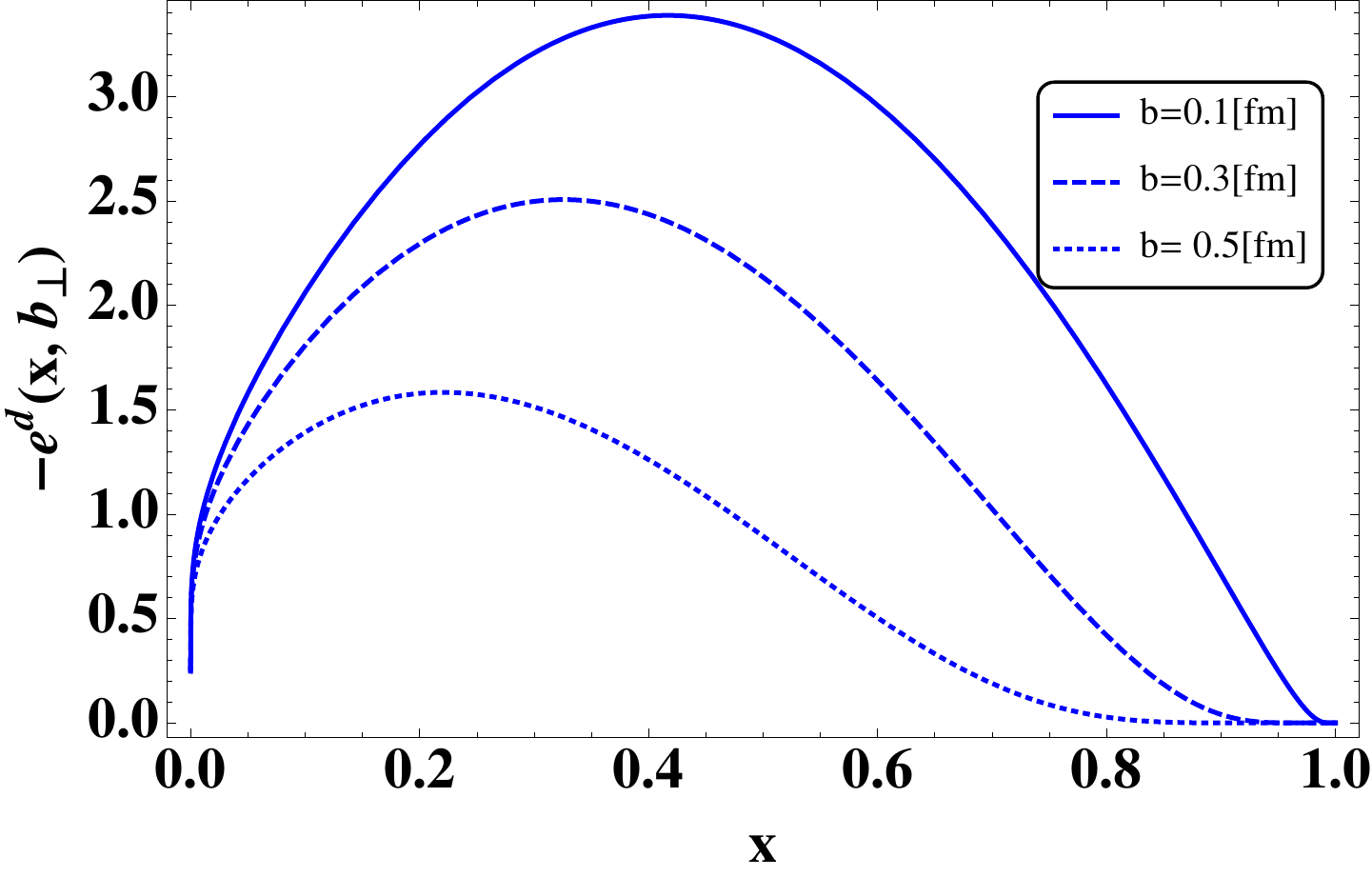}
\hspace{0.1cm}
\small{(d)}\includegraphics[width=7cm,height=5cm,clip]{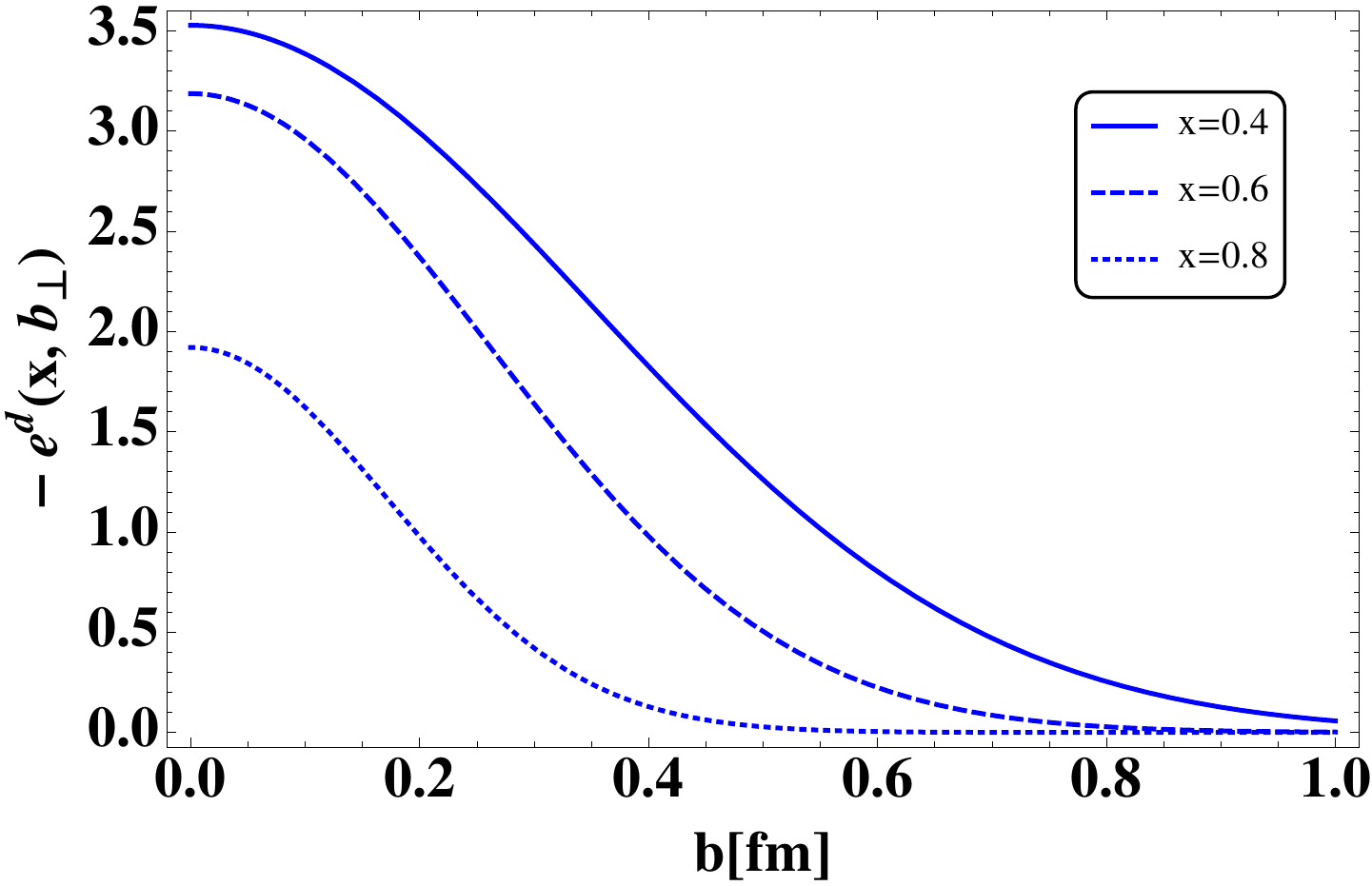}
\end{minipage}
\caption{\label{eub}(Color online) Plots of (a)  $e^u(x,b_\perp)$ vs $x$ for fixed values of $b=0.1, 0.3, 0.5$ fm for up quark;  (b)  $e^u(x,b_\perp)$ vs the impact parameter $b = |b_\perp|$ for fixed value  of $x=0.4,0.6,0.8$ for up quark; (c) $-e^d(x,b_\perp)$ vs $x$ for fixed values of $b$ for down quark; and (d) $-e^d(x,b_\perp)$ vs $b$ for fixed value  of $x$ for down quark.}
\end{figure}

In Fig. \ref{hub}(a), we have plotted the behavior of $u(x,b)$ with $x$ for fixed values of $b=0.1, 0.3, 0.5$ fm and in Fig. \ref{hub}(b), we have shown the behavior of same GPD  with the impact parameter $b$ for fixed values of $x =0.4, 0.6, 0.8$. In Figs. \ref{hub}(c)$-$\ref{hub}(d), we plot the same GPDs $d(x,b)$ for the down quark for the same set of parameters.  Similar plots showing the behavior of  GPDs $e^{u/d} (x, b)$ are shown in Fig. \ref{eub}. The qualitative behavior of  GPDs $q(x, b)$ and $e^{q} (x, b)$ is the same for both up and down quarks.  In both cases, the maxima of GPDs shifted towards a lower value of $x$ as $b$ increases; therefore the transverse profile is peaked at $b=0$ and falls off further.  It is also interesting to observe that for the small values of $b$, the magnitude of GPD $q(x,b)$ is larger for the up quark than down quark, whereas the magnitude of the GPD $e^q(x,b)$ is marginally larger for the down quark than up quark.

An estimate of the transverse size of a hadron is given by the generalized transverse mean squared radius \cite{mlattice2}
\be  {\langle r^{2n}_\perp  \rangle}^q_{ch}  = { \int {\mathrm d^2}  b_\perp \, b^2_\perp  \int{\mathrm d}  x x^{n-1} q(x, b_\perp) \over \int {\mathrm d^2}  b_\perp \,  \int{\mathrm d}  x x^{n-1} q(x, b_\perp) }  
= { \int {\mathrm d^2}  b_\perp \, b^2_\perp  H^q_n (Q^2)  \over  \int {\mathrm d^2}  b_\perp \,  H^q_n (Q^2) }\,, \ee
which coincides with the standard transverse mean square charge radius for $n=1$.  The transverse charge radii in the modified soft-wall model  for the up and down quarks are comparable with each other ($ {\langle r^{2}_\perp  \rangle}^u_{ch} = 0.406 ~$fm$^2$  and ${\langle r^{2}_\perp  \rangle}^d_{ch}= 0.372$~fm$^2$).  The transverse size of the nucleon depends significantly on parameter $x$; we define the transverse width
 \be {\langle r^2_\perp (x)  \rangle}^q_{ch} = { \int {\mathrm d^2}  b_\perp \, b^2_\perp  q(x, b_\perp) \over 
 \int {\mathrm d^2}  b_\perp \,  q(x, b_\perp) }  = -4 {\partial \log H(x,Q^2)  \over  \partial Q^2}  |_{Q^2=0}  =  { \log (1/x) \over \kappa^2 }\,. \ee
It is significant to note here that the $x$-dependence of the transverse width in the modified soft-wall model is the same as the soft-wall model \cite{soft}.

\section{Transverse charge and magnetization densities}
\label{mag}

The  charge and magnetization densities in the transverse plane for an unpolarized nucleon are defined by the Fourier transforms of EFFs. Since the  EFFs are related to the GPDs via the sum rules,  the transverse charge densities are  related to the impact parameter dependent GPDs. The quark transverse densities can be obtained from the flavor decomposition of the densities of the proton and neutron by invoking the charge and isospin symmetry. The density functions are not directly measured in experiments; however, an estimate can be obtained from  the  analysis of hard-scattering data on EFFs \cite{kelly}. We compare our results with the global fit to data on EFFs  using  the functional form of $Q^2$ dependence, referred  to as ``Kelly parametrization''. The  charge density $(\rho_{ch}^N)$  in transverse impact parameter space is expressed as
 \be
\rho_{ch}^N (b_\perp) = \sum_q e_q^N \int_0^1 {\mathrm d} x\, q(x, b_{\perp})=
  \int { {\mathrm d^2} q_\perp \,  \over (2 \pi)^2 }  F_1^N(q^2) e^{\iota q_\perp \cdot b_\perp}  \,. \label{ch} \\ 
\ee
In the modified soft-wall model of AdS/QCD, the results for  transverse charge density for an unpolarized nucleon are
\bea
\rho_{ch}^N (b_\perp) &=& {\kappa^2 \over \pi} \sum_q e_q^N  \sum\limits_\tau \, c_\tau \, \int^1_0 {\mathrm d}x \, {q(x, \tau)\over \log(1/x)}  e^{b_\perp^2 \kappa^2 \over \log(x)}  \label{ch1} \,.\eea

\begin{figure}[htbp]
\begin{minipage}[c]{0.98\textwidth}
\small{(a)}
\includegraphics[width=7cm,height=5cm,clip]{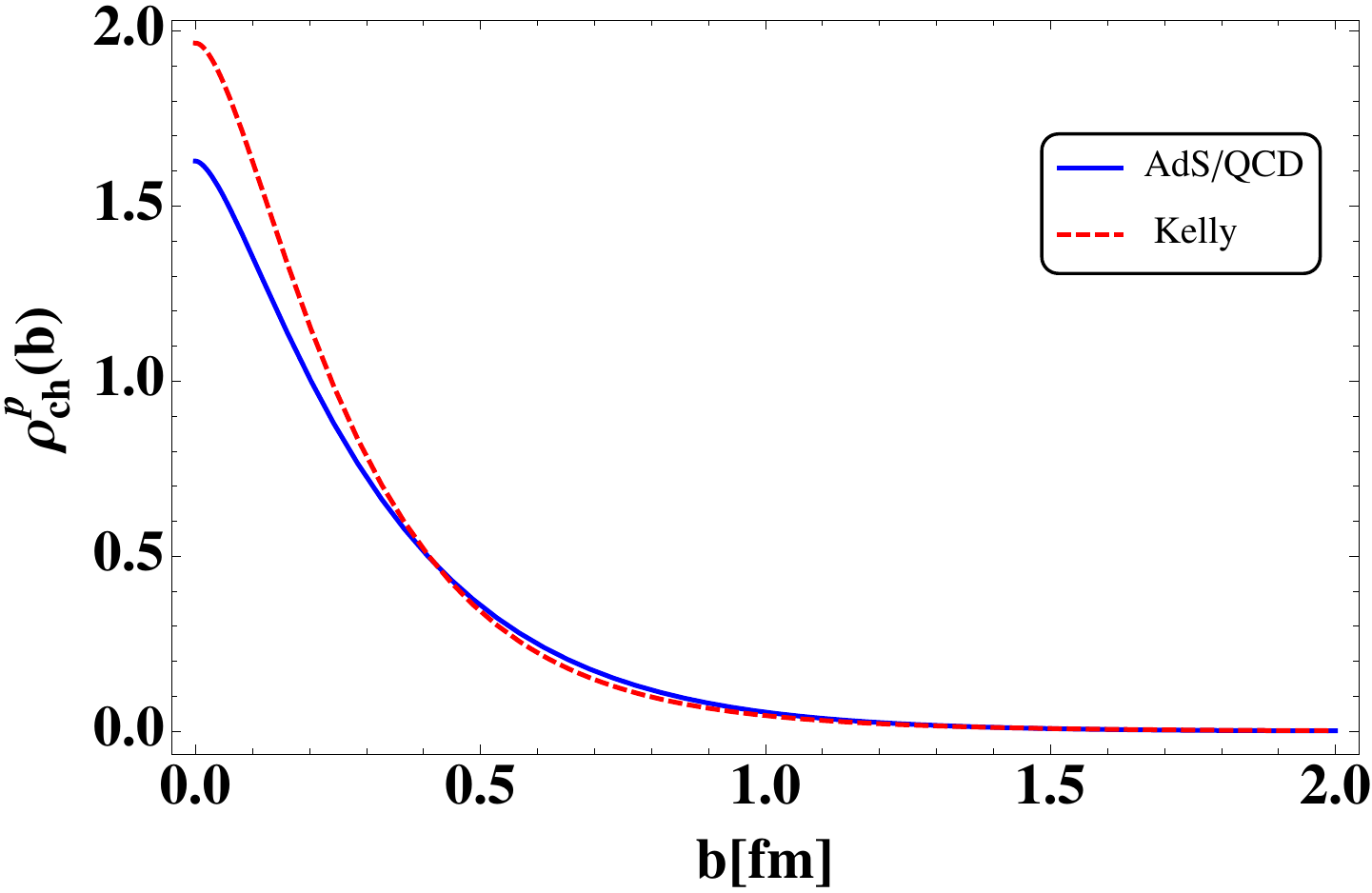}
\hspace{0.1cm}
\small{(b)}\includegraphics[width=7cm,height=5cm,clip]{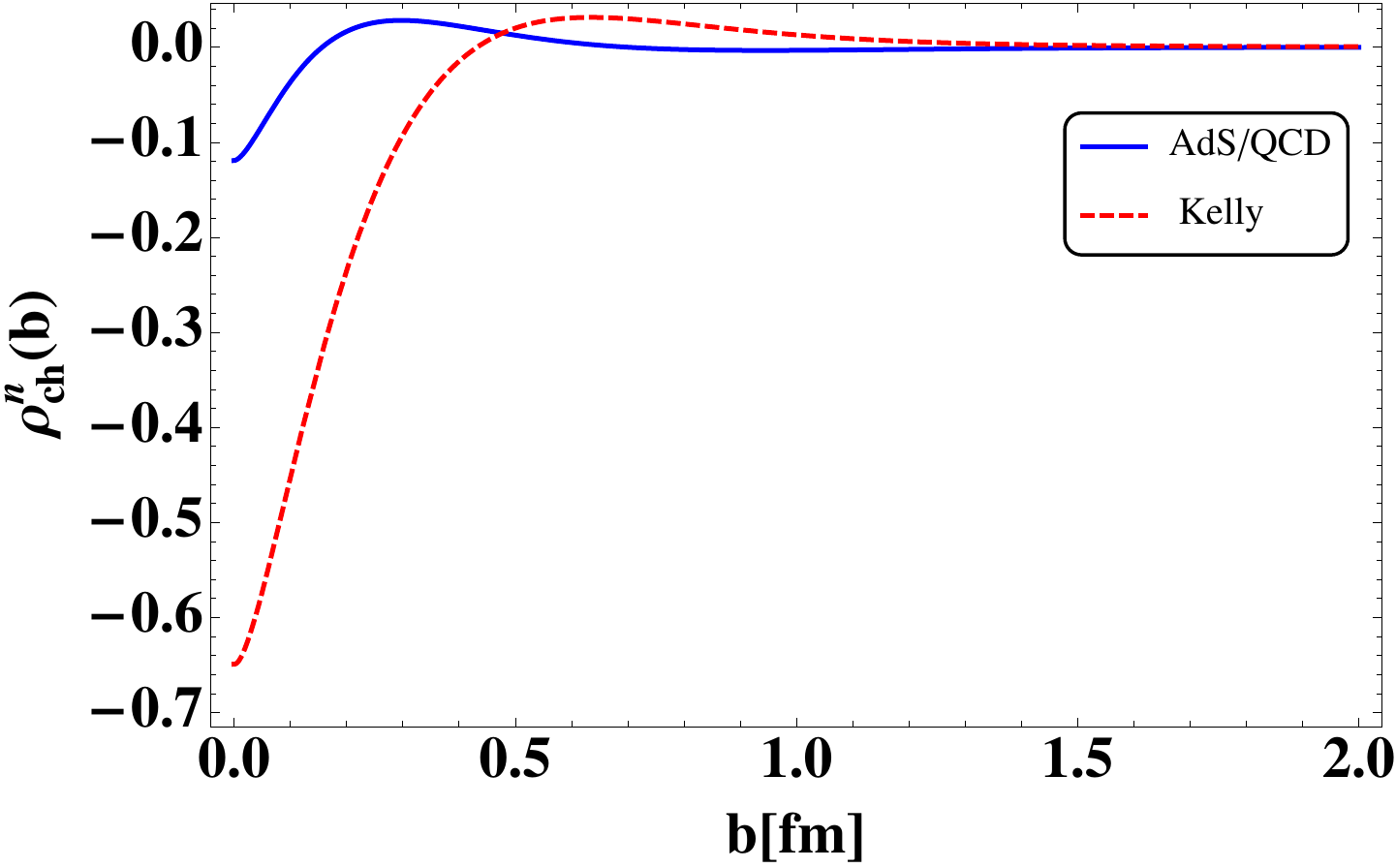}
\end{minipage}
\begin{minipage}[c]{0.98\textwidth}
\small{(c)}\includegraphics[width=7cm,height=5cm,clip]{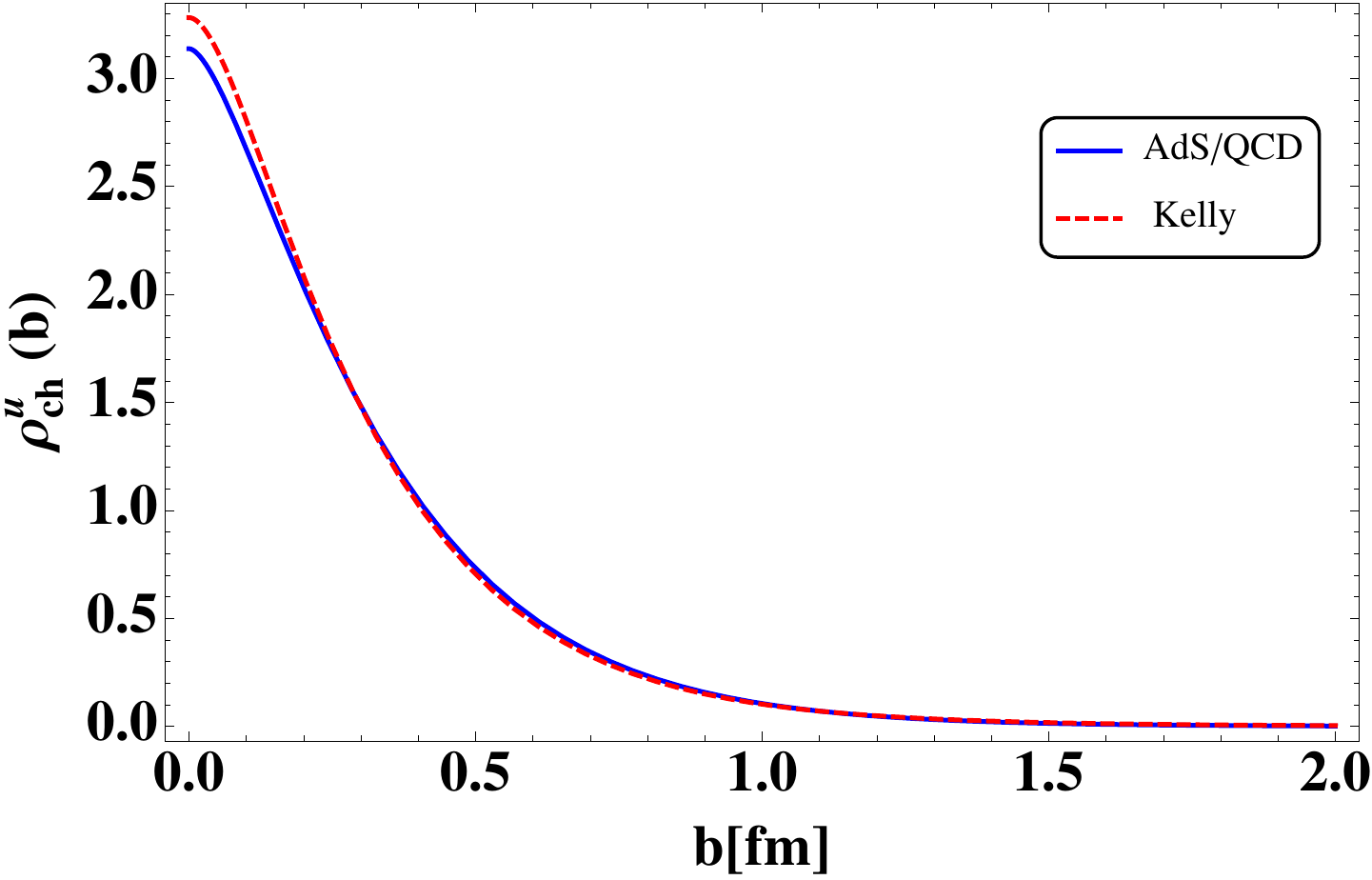}
\hspace{0.1cm}
\small{(d)}\includegraphics[width=7cm,height=5cm,clip]{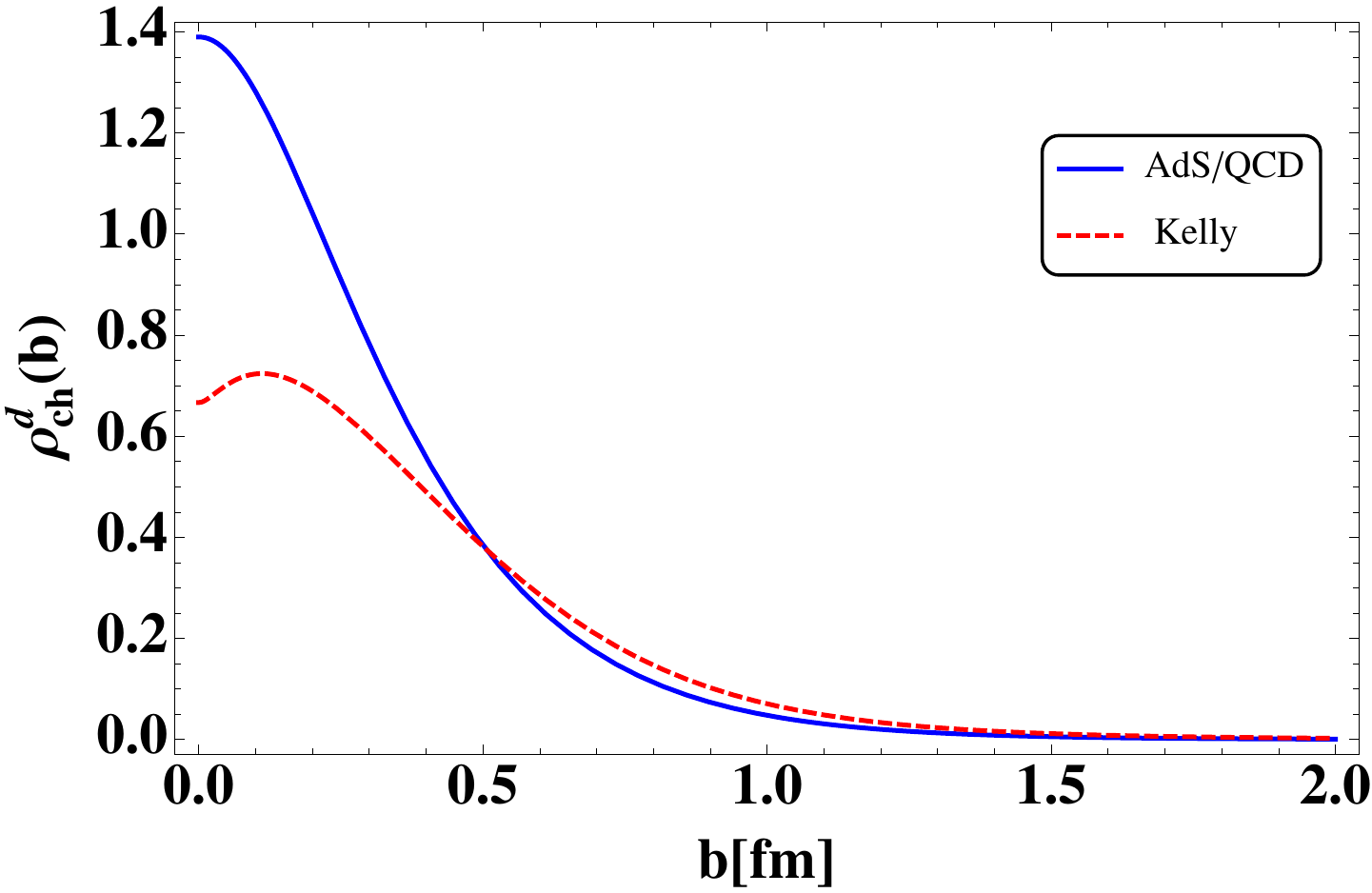}
\end{minipage}
\caption{\label{rep}Plots of transverse charge  densities $\rho_{ch}^N(b_\perp)$  with impact parameter $b_\perp$ for (a) proton  and  (b) neutron. (c) and (d) represent the corresponding contributions from the up and down quarks. The solid line represents the  predictions of the AdS/QCD and dashed line correspond to the Kelly parametrization \cite{kelly}.
}
\end{figure}

In Figs.  \ref{rep}(a)$-$\ref{rep}(b), we have plotted  the transverse charge density with the impact parameter $b$ for proton and neutron in the AdS/QCD and Kelly parametrization.   The overall trend in the behavior of densities is the same as Kelly parametrization \cite{miller2007}.  It has been observed that the proton charge density has a large positive value at the center of the core which falls off  further as $b$ increases. The neutron charge density reveals a negative core at $b$ values smaller thqn 0.3 fm, a positive contribution at the intermediate $b$ values, and negative contribution at large distances more 1.5 fm.  The results for the neutron charge density are more interesting as they contradict  the previous idea of positive charge density at core due to one gluon exchange or pion-cloud contribution.  Our results match well with the findings of a model-independent analysis of charge density of the neutron  using the data on  EFFs \cite{miller2007}.

To get an insight into the contributions of the different quark flavors, we have plotted the charge densities for the up and down quarks in Figs. \ref{rep}(c)$-$\ref{rep}(d).  One can observe  that the up quark charge density is large and positive  when compared to the down quark this leads to the positive value of proton charge density over the entire range. On the other hand, the contribution of both up and down quarks are comparable for the  neutron.  At the center of the neutron the up quarks have more negative charge density which gives the negative core surrounded by  the positively charged proton.  Future experimental  information on neutron EFFs  could  render the  present situation more precise. 

The  magnetization density $(\rho_m^N)$  in transverse impact parameter space is given as
 \be
{\tilde \rho_{m}^N (b_\perp)} =  \sum_q e_q^N \int_0^1 {\mathrm d} x\, e^q(x, b_{\perp})
=\int { {\mathrm d^2} q_\perp \,  \over (2 \pi)^2 }  e^{\iota q_\perp \cdot b_\perp}   F^N_2(q^2)   \,,
\ee
however, the true anomalous magnetization density in the transverse plane is expressed as
\be  \rho_m^N(b_\perp) = -b{ \partial  {\tilde \rho_m^N (b)} \over  \partial b} \,, \ee
In  modified soft-wall model of AdS/QCD, we have \bea  \rho_m^N (b_\perp) &=& {2 b^2  \kappa^4 \over \pi }  \sum_q e_q^N  \sum\limits_\tau \, c_\tau \, \int^1_0 {\mathrm d}x \, { {e^q}(x, \tau) \over {\log(1/x)}^2 } e^{b_\perp^2 \kappa^2 \over \log(x)} \,.  \label{amd}
\eea

\begin{figure}[htbp]
\begin{minipage}[c]{0.98\textwidth}
\small{(a)}
\includegraphics[width=7cm,height=5cm,clip]{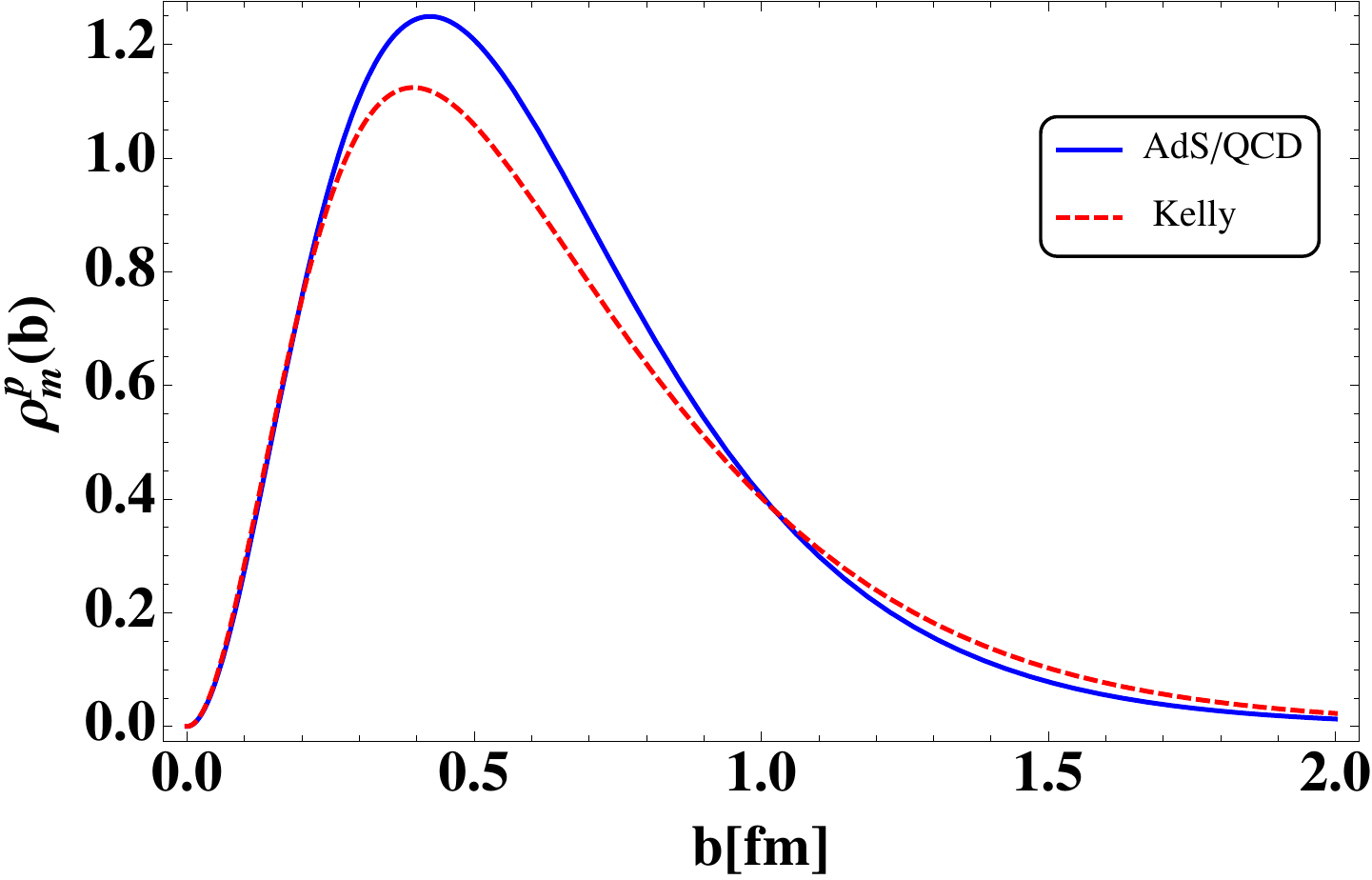}
\hspace{0.1cm}
\small{(b)}\includegraphics[width=7cm,height=5cm,clip]{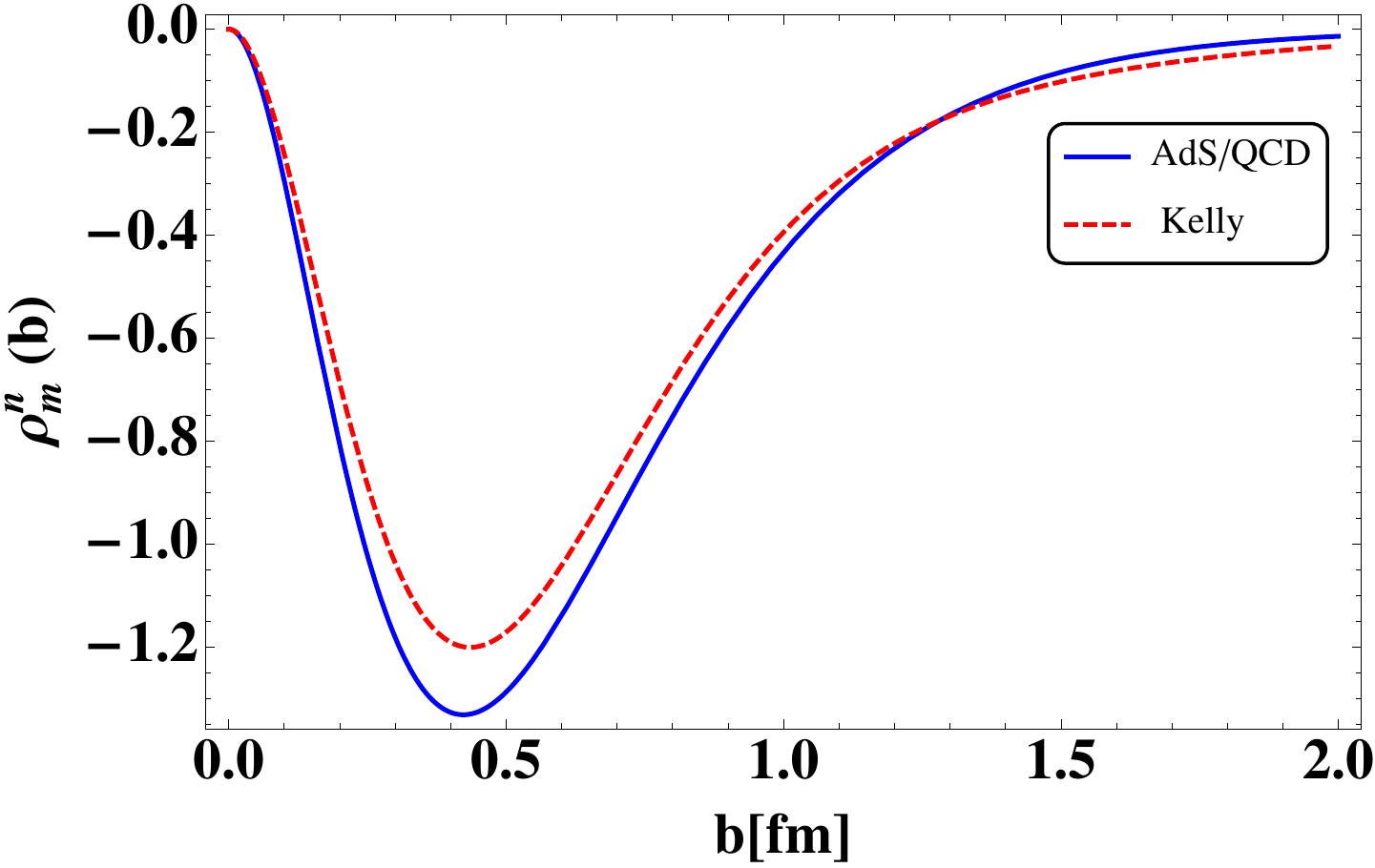}
\end{minipage}
\begin{minipage}[c]{0.98\textwidth}
\small{(c)}\includegraphics[width=7cm,height=5cm,clip]{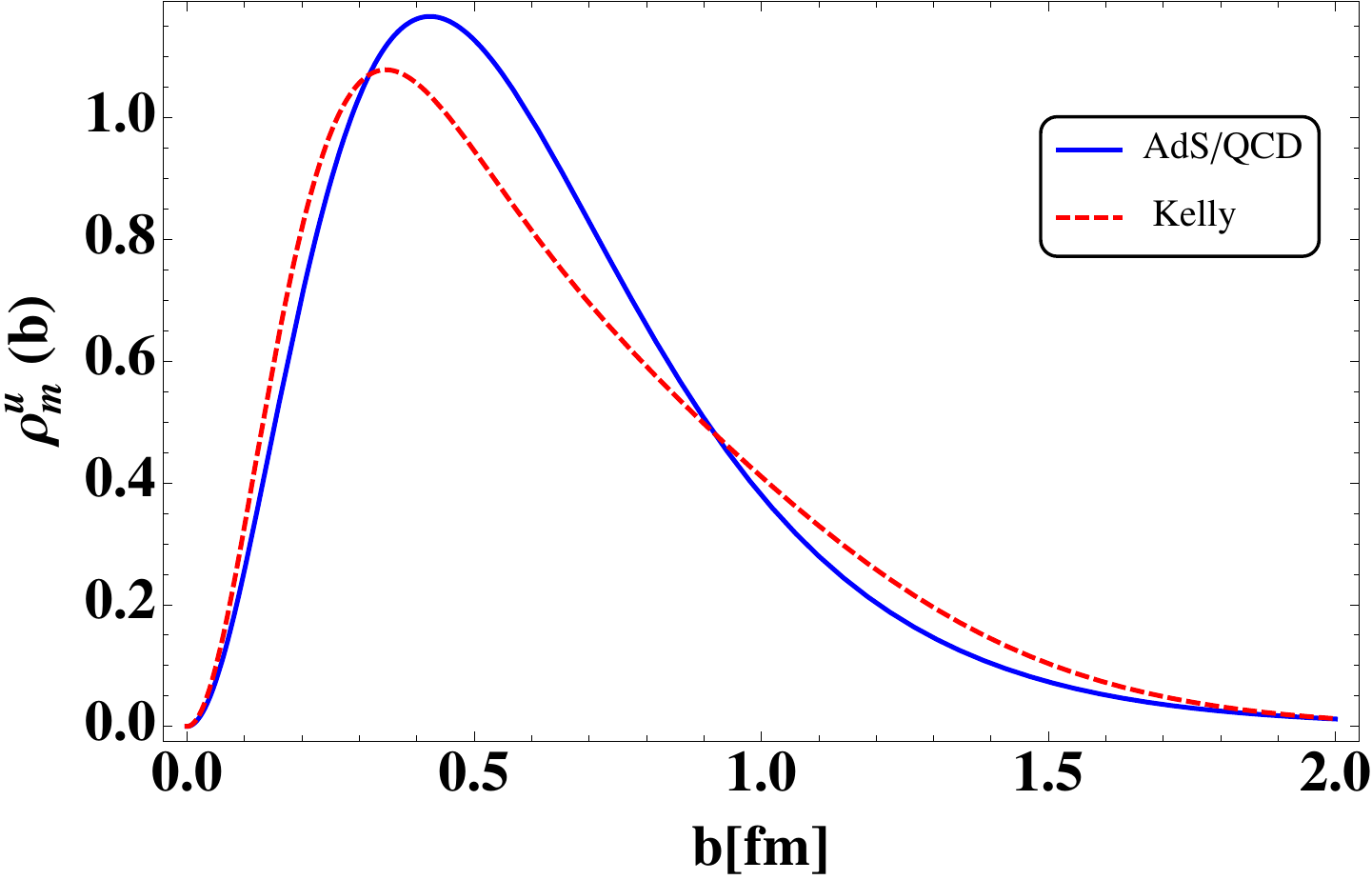}
\hspace{0.1cm}
\small{(d)}\includegraphics[width=7cm,height=5cm,clip]{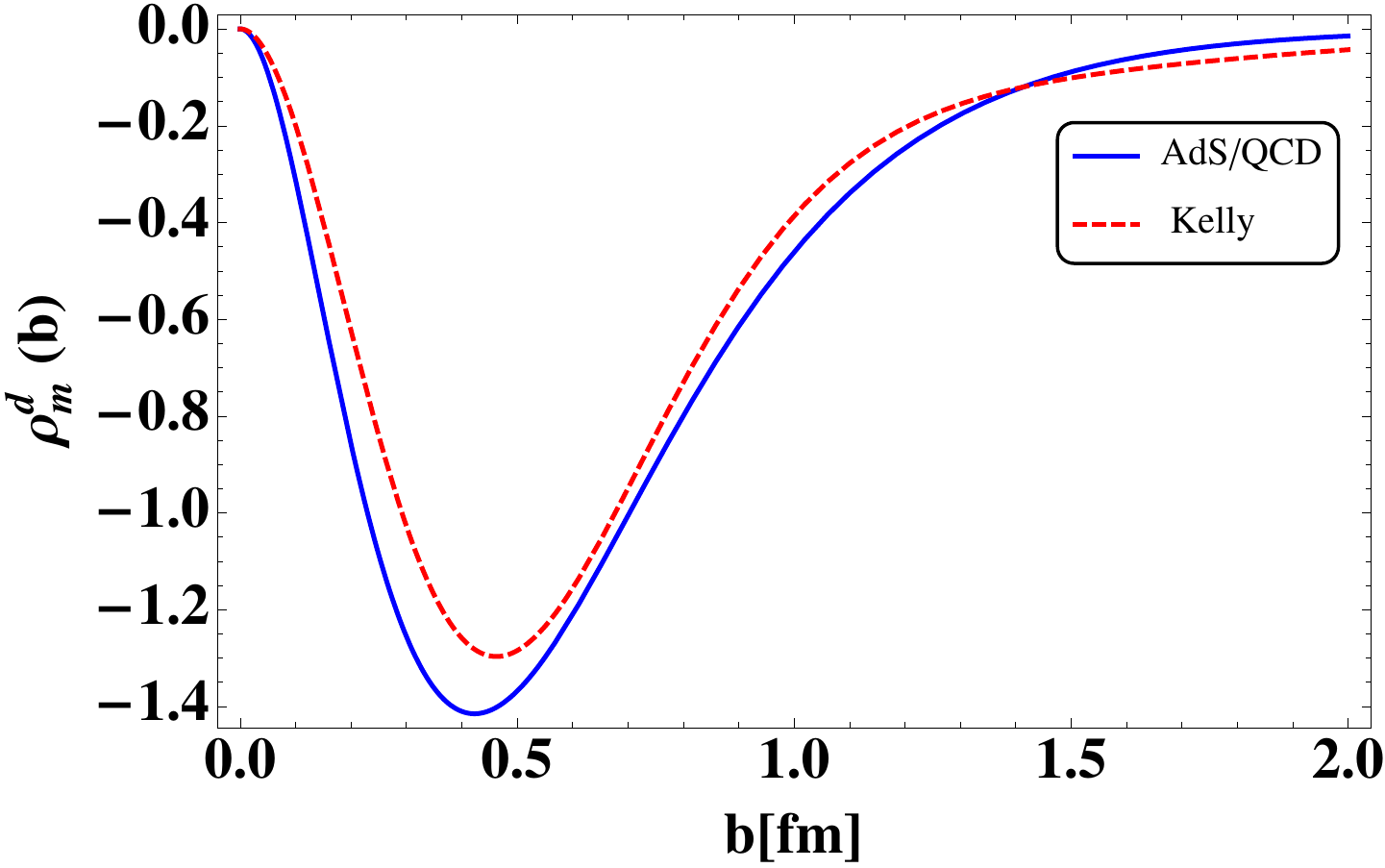}
\end{minipage}
\caption{\label{ramp}Plots of transverse anomalous magnetization densities $\rho_m^N (b_\perp)$  with impact parameter  $b_\perp$ for (a) proton  and  (b) neutron. (c) and (d) represent the contributions from the up and down quarks. The solid line represents the  predictions of the AdS/QCD and dashed line correspond to the  Kelly parametrization \cite{kelly}.
}
\end{figure}

In Figs.  \ref{ramp}(a)$-$\ref{ramp}(d), we have plotted the behavior of anomalous magnetization densities with the impact parameter $b$ for proton and neutron and their flavor contribution in the AdS/QCD holography and Kelly parametrization. It is interesting to observe that the our model predictions overlap significantly with the predictions of Kelly parametrization for  all case.  The transverse anomalous magnetization density is positive  for the proton and negative for  neutron in consistency with  the measured values of the anomalous magnetic moments. The magnetization densities of the up and down quarks are comparable but both are magnetized in opposite direction.

%%%%%%%%%%%%%%%%%%%

It is also instructive to investigate the transverse charge densities inside the transversely polarized nucleon. We consider a nucleon polarized in the $xy$ direction with the transverse polarization direction $S_\perp= \cos \phi_s {\hat x} + \sin \phi_s {\hat y}$.  Following the Ref. \cite{carl}, charge density in transverse plane for  transversely polarized nucleon 
\bea \rho_T^N (b) &=&  \rho_{ch}^N (b) - {\sin (\phi_b - \phi_s)   \over 2 M_N b}  \rho_{m}^N (b)  \,, \label{rhoT} 
  \eea
where $M_N$ is mass of nucleon and the transverse impact parameter $b_\perp= b( \cos \phi_b {\hat x} + \sin \phi _b {\hat y})$.  The second term in the above expression measure the deviation from circular symmetric unpolarized charge density and  depends on the orientation of $b_\perp$ relative to the transverse spin direction $S_\perp$.

\begin{figure}[htbp]
\begin{minipage}[c]{0.98\textwidth}
\small{(a)}
\includegraphics[width=7cm,height=7cm,clip]{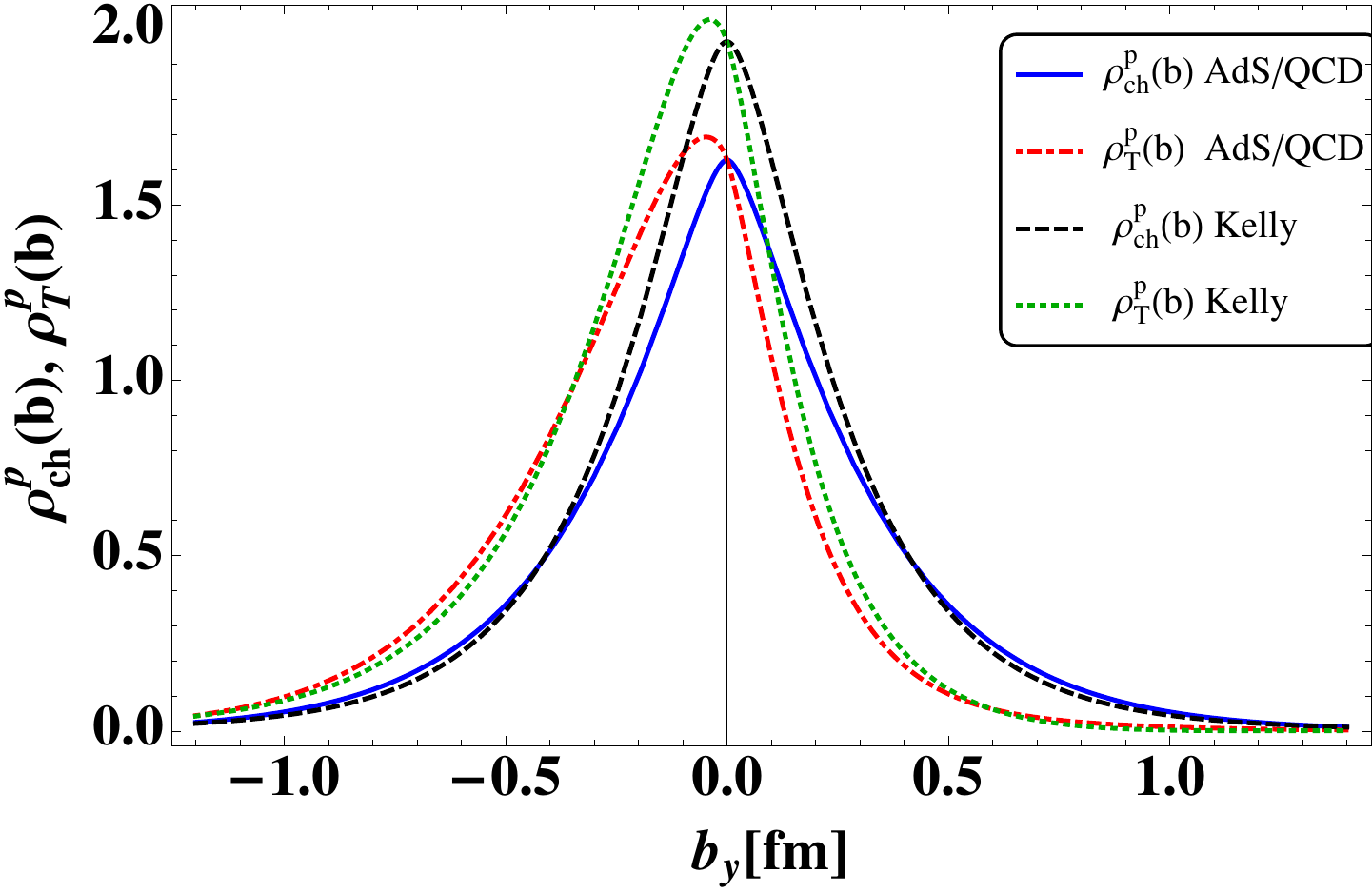}
\hspace{0.1cm}
\small{(b)}\includegraphics[width=7cm,height=7cm,clip]{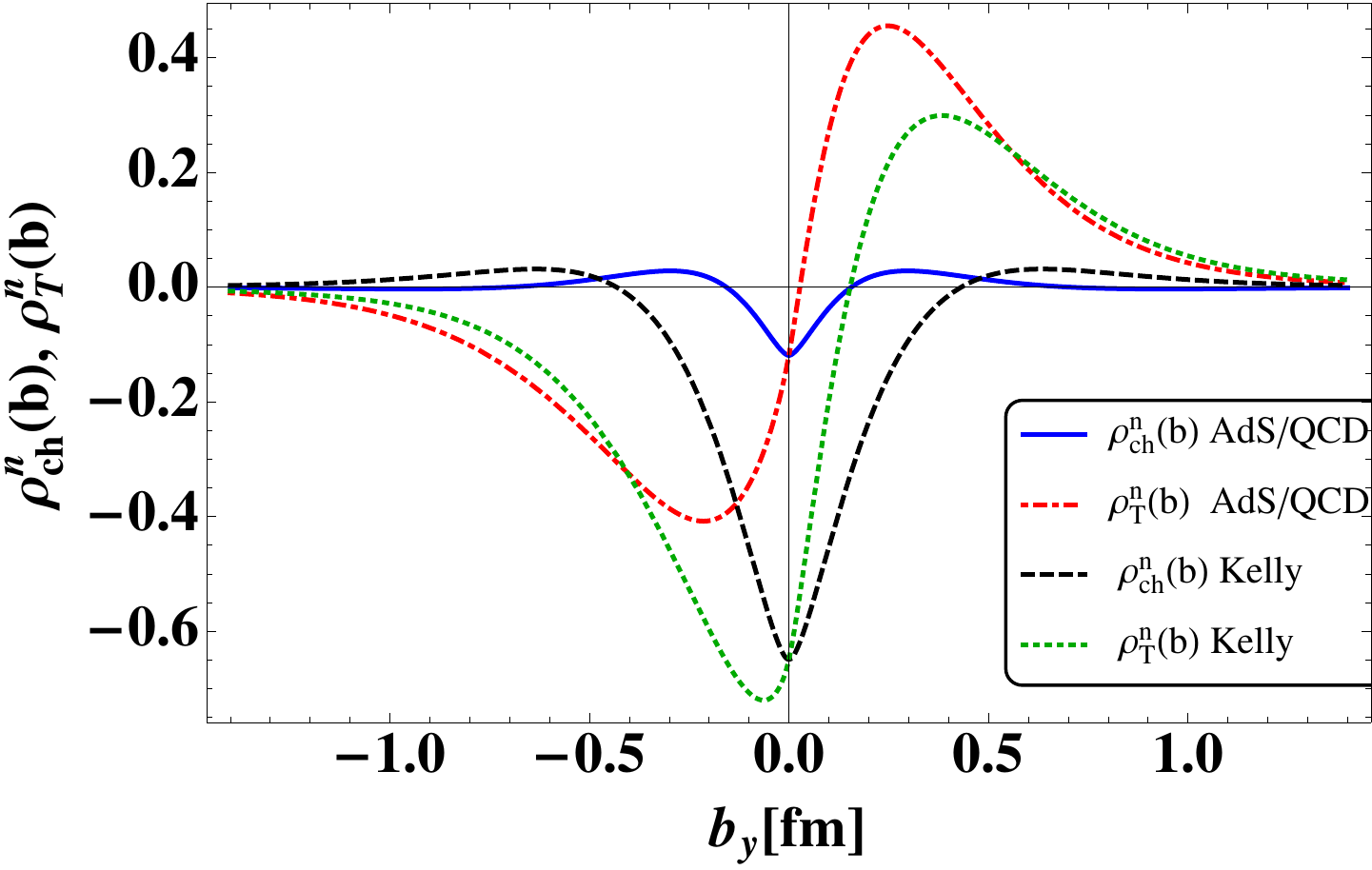}
\end{minipage}
\begin{minipage}[c]{0.98\textwidth}
\small{(c)}\includegraphics[width=7cm,height=7cm,clip]{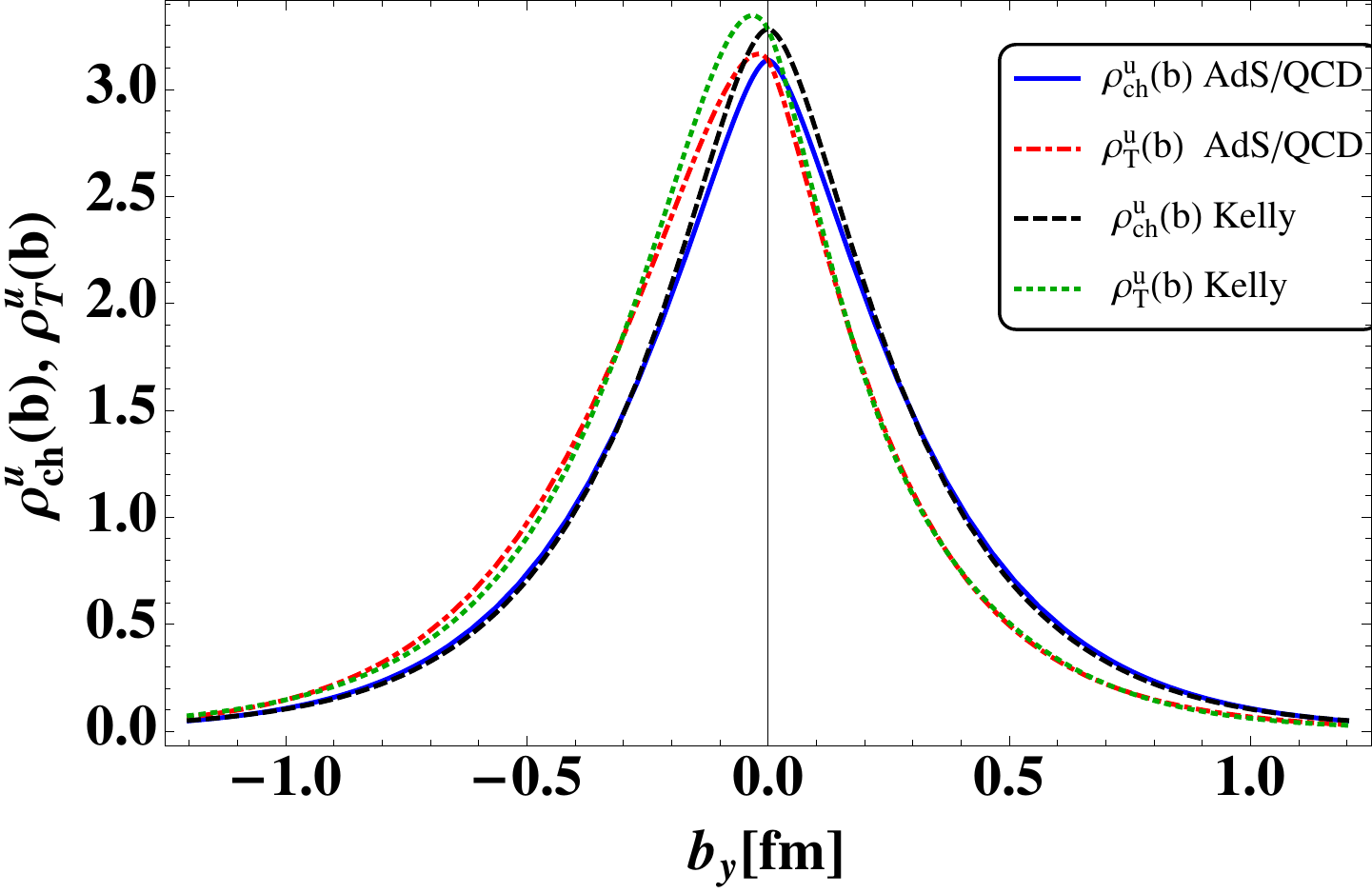}
\hspace{0.1cm}
\small{(d)}\includegraphics[width=7cm,height=7cm,clip]{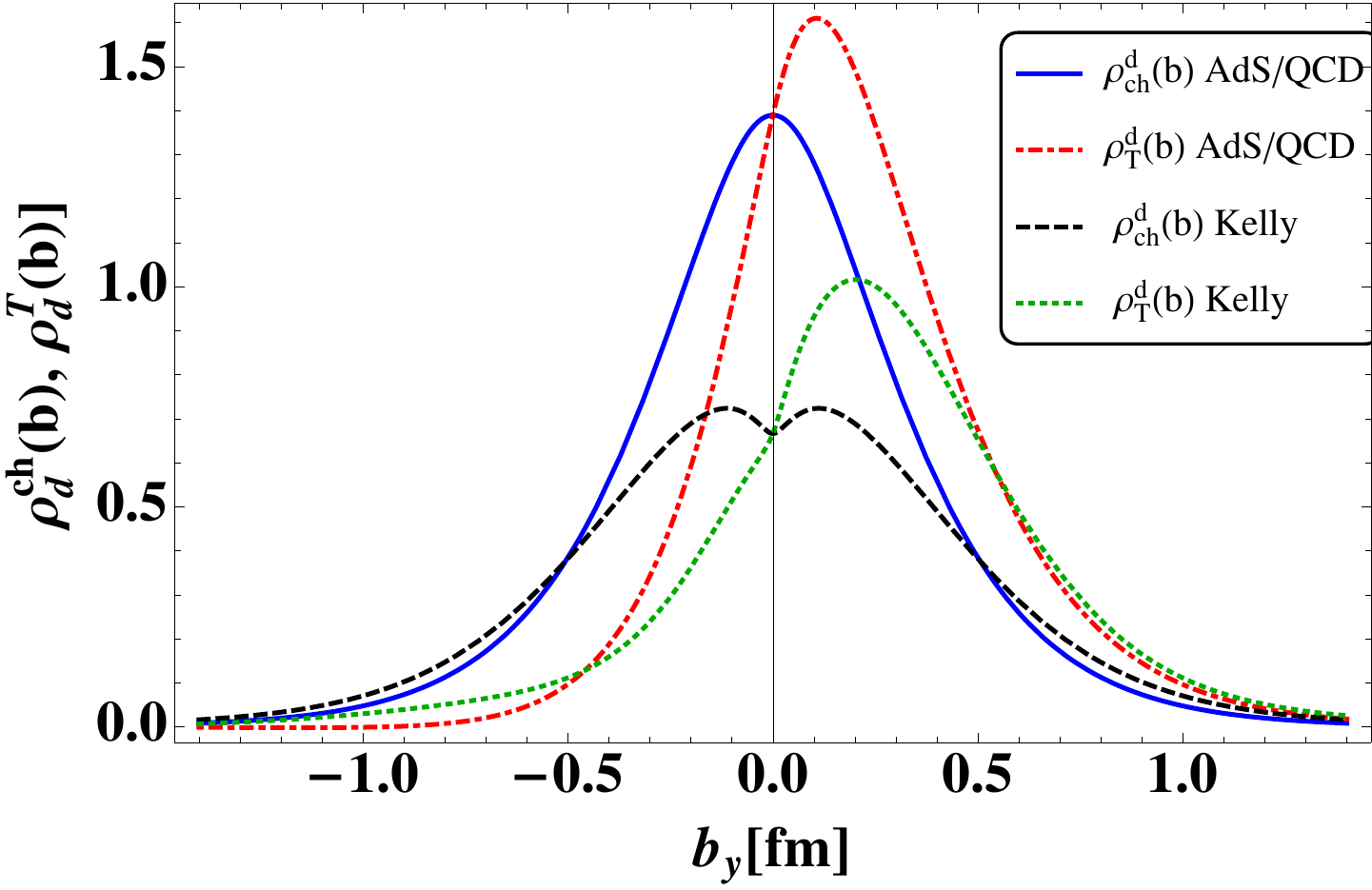}
\end{minipage}
\caption{\label{cmp}Plots of transverse charge density $\rho_{ch}^N (b)$ and transverse anomalous magnetization densities $\rho_m^N (b)$  with transverse impact parameter space $b_\perp$ for (a) proton  and  (b) neutron, (c) and (d) represent the contributions from the up and down quarks. The solid line represents the  predictions of the AdS/QCD and dashed line represents the  Kelly parametrization \cite{kelly}. }
\end{figure}

In Figs. \ref{cmp}(a)$-$\ref{cmp}(b), we present a comparison of the behavior of charge density for the unpolarized and transversely polarized proton and neutron and compare them with the Kelly parametrization.   The transverse charge densities for the proton polarized transversely along the positive $x$ direction ($\phi_s=0$) are distorted in the negative $b_y$ direction. The transverse polarization of the proton leads to an induced electric dipole moment along the negative $b_y$ axis due to relativistic effects. In  the case of the neutron, the negative charge is located in the center of the neutron surrounded by the positive charge. When the neutron is transversely polarized along the $x$ axis, the negative charge is shifted to the negative $b_y$ direction and the positive charge move towards to the positive $b_y$ direction. This result follows from the fact that the neutron anomalous magnetic moment itself is negative, which yields an induced electric dipole moment along the positive $b_y$ axis. Our results are the same as the predictions of the other phenomenological models, such as the chiral quark soliton model \cite{csqm}, lattice QCD \cite{lattchargeden},  finite radius approximation \cite{venkat},  parametrization approach \cite{carl}, soft-wall model  \cite{dipt}, etc.

\begin{figure}[htbp]
\begin{minipage}[c]{0.98\textwidth}
\small{(a)}
\includegraphics[width=7cm,height=6cm,clip]{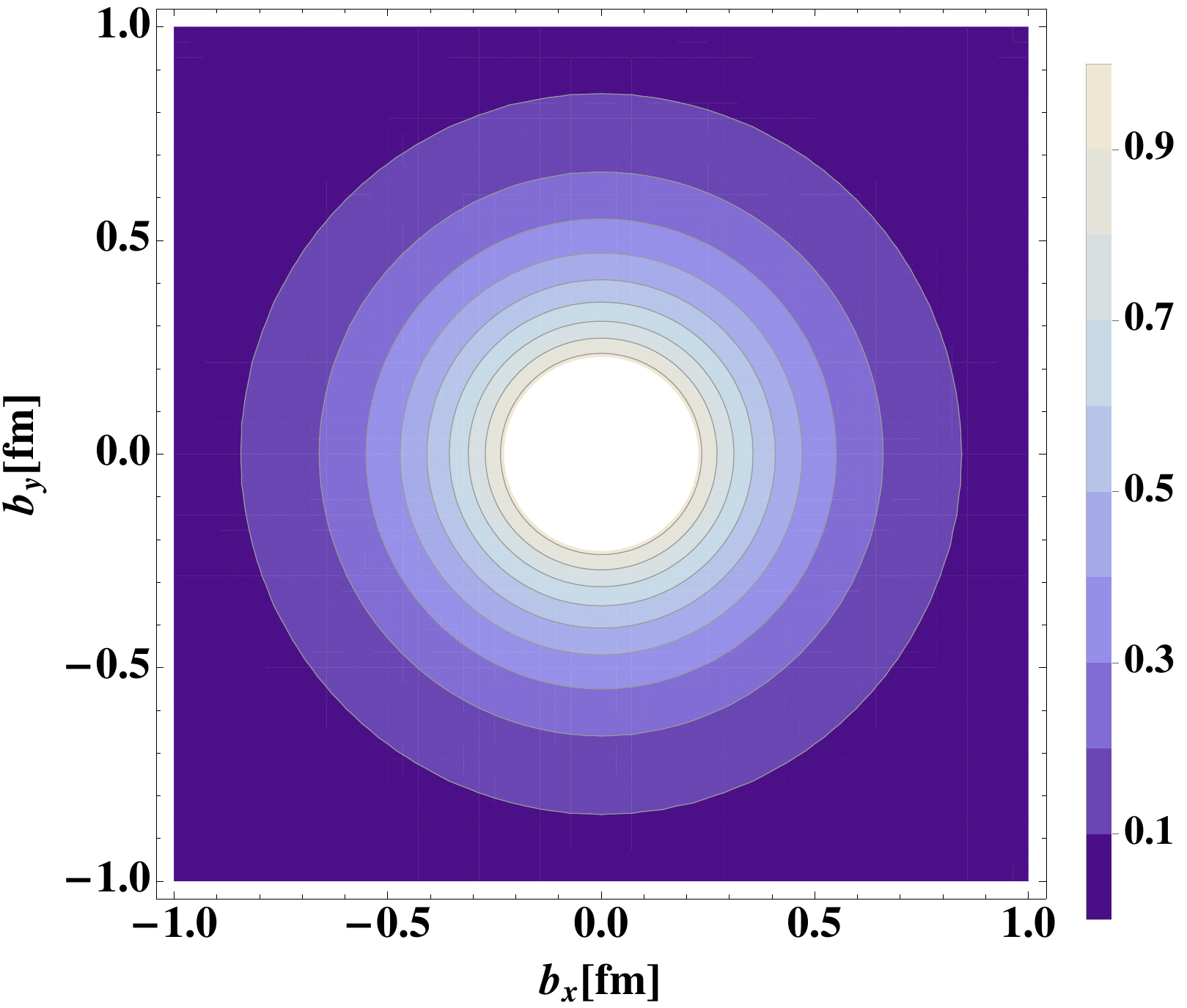}
\hspace{0.1cm}
\small{(b)}\includegraphics[width=7cm,height=6cm,clip]{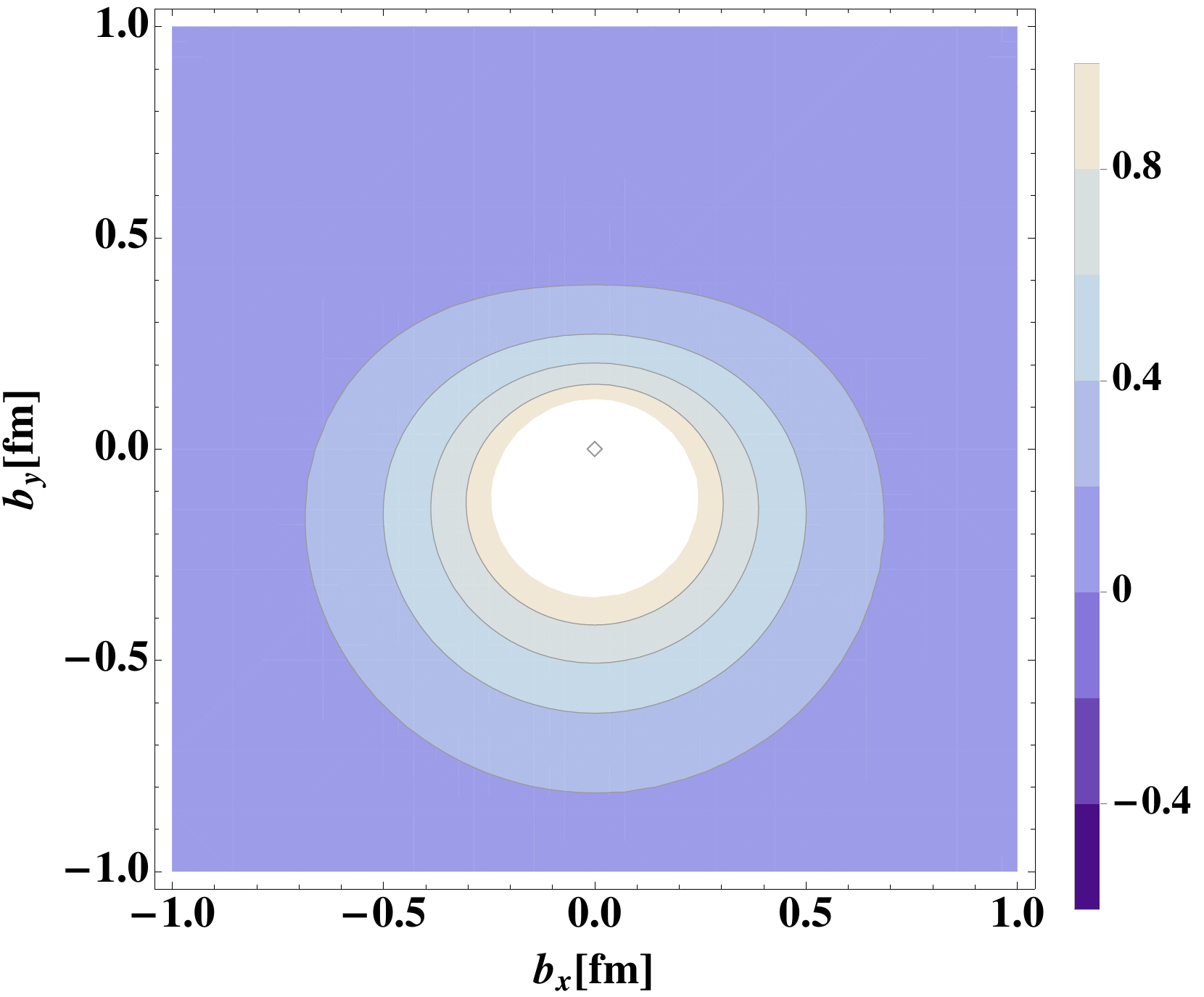}
\end{minipage}
\begin{minipage}[c]{0.98\textwidth}
\small{(c)}\includegraphics[width=7cm,height=6cm,clip]{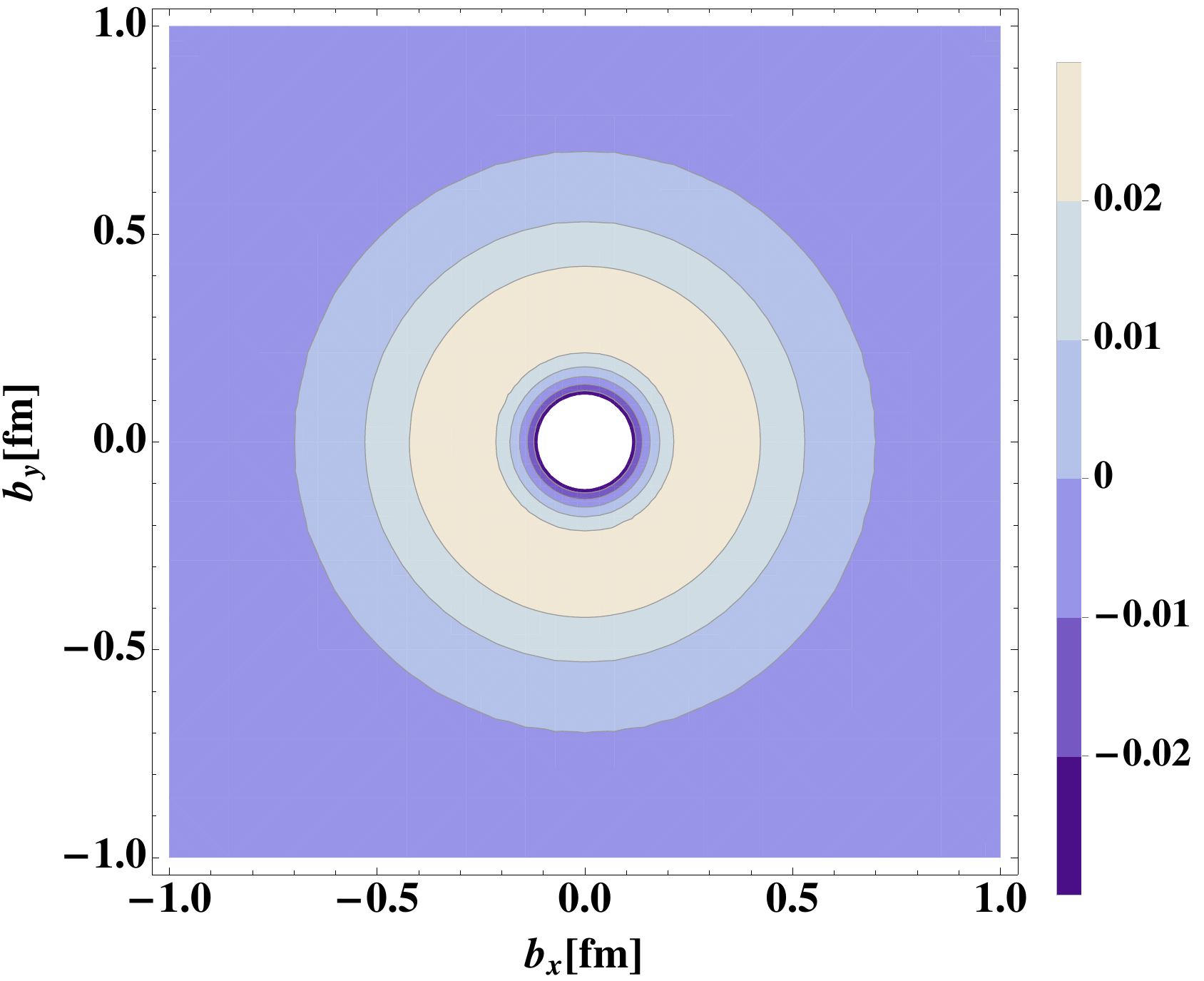}
\hspace{0.1cm}
\small{(d)}\includegraphics[width=7cm,height=6cm,clip]{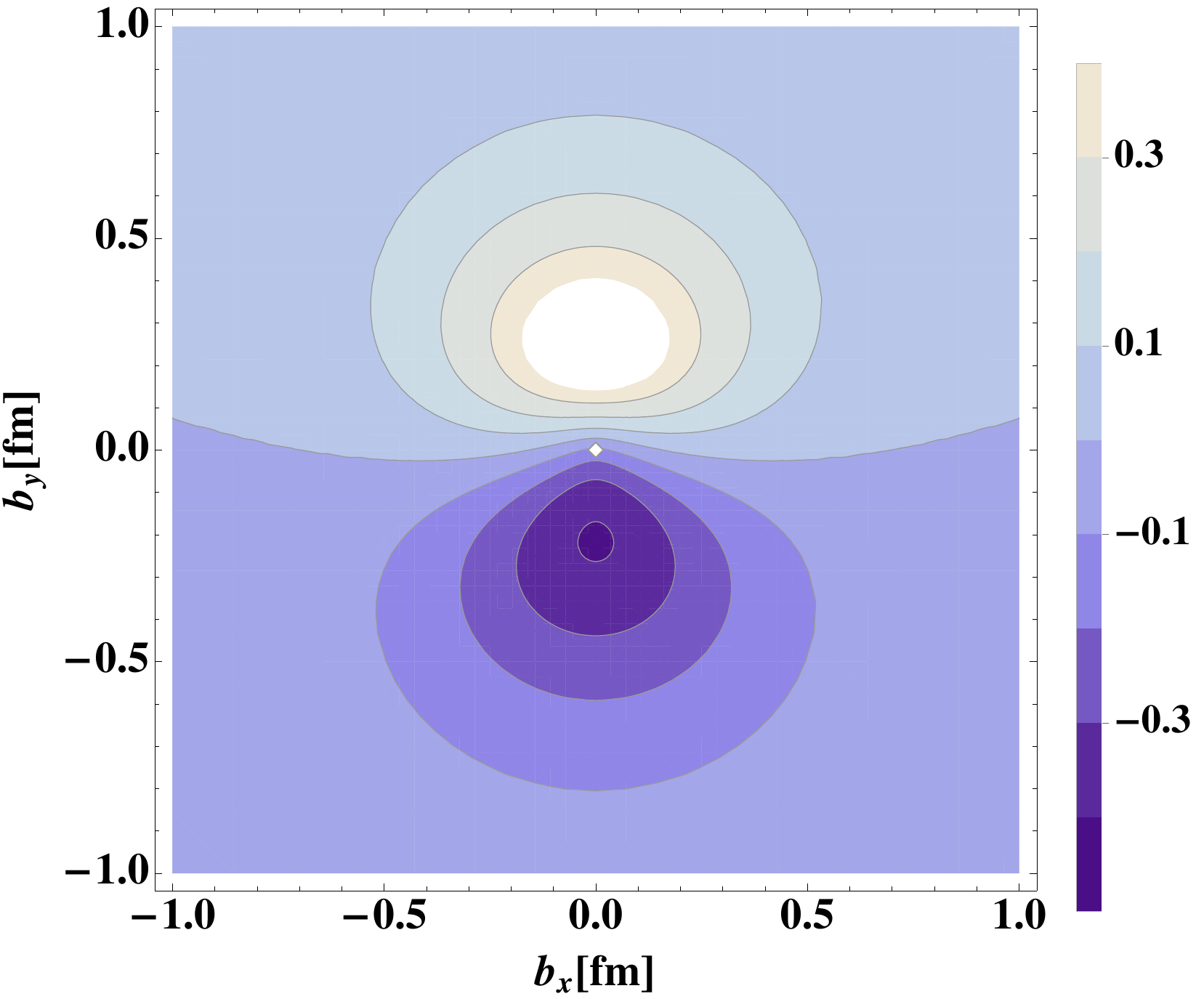}
\end{minipage}
\caption{\label{bcp}Plots of charge densities in the transverse plane for the (a) unpolarized proton, (b) transversely polarized proton,  (c) unpolarized neutron, and (d) transversely polarized neutron. The transverse polarization is taken along the $x$ direction. }
\end{figure}

\begin{figure}[htbp]
\begin{minipage}[c]{0.98\textwidth}
\small{(a)}
\includegraphics[width=7cm,height=6cm,clip]{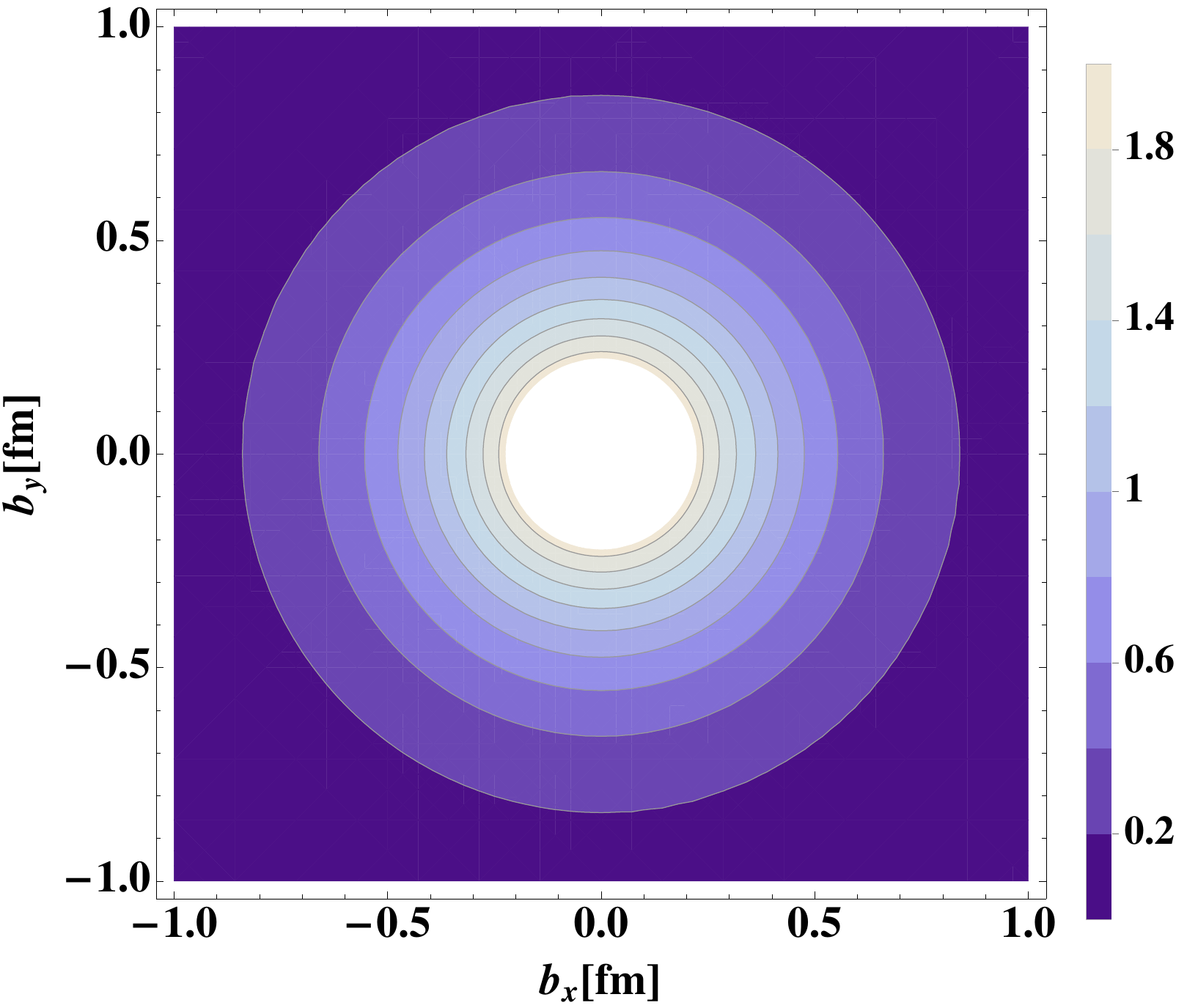}
\hspace{0.1cm}
\small{(b)}\includegraphics[width=7cm,height=6cm,clip]{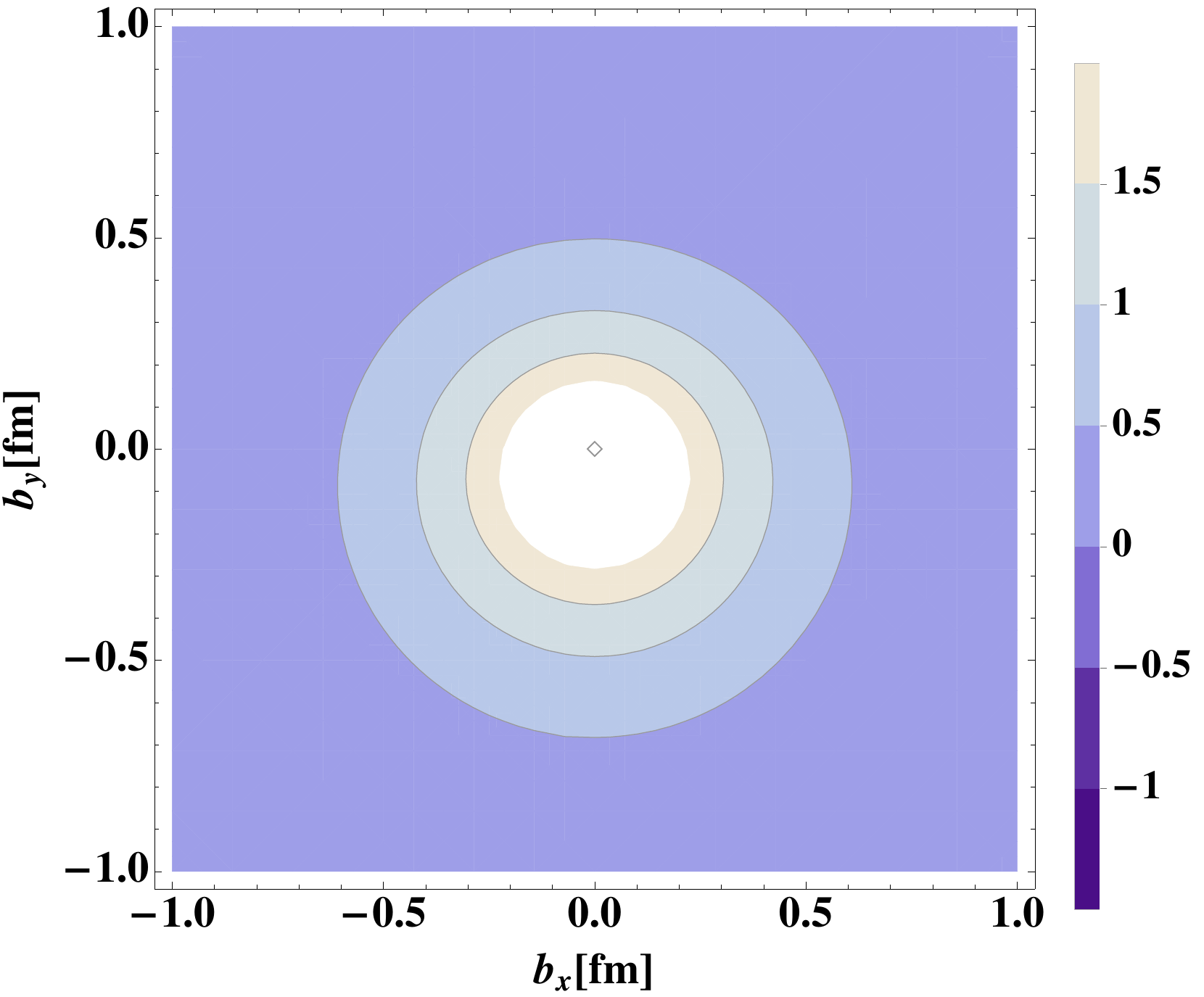}
\end{minipage}
\begin{minipage}[c]{0.98\textwidth}
\small{(c)}\includegraphics[width=7cm,height=6cm,clip]{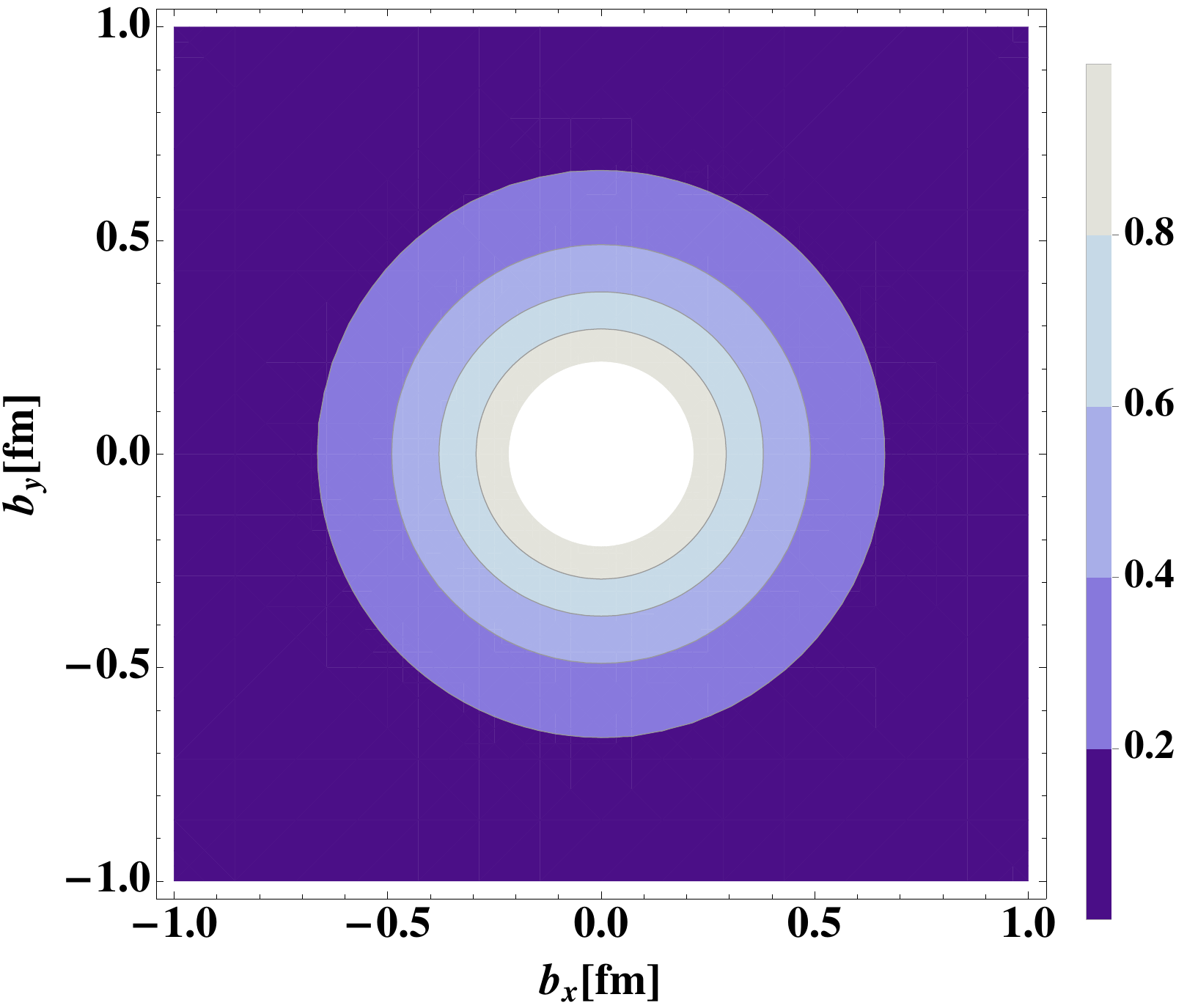}
\hspace{0.1cm}
\small{(d)}\includegraphics[width=7cm,height=6cm,clip]{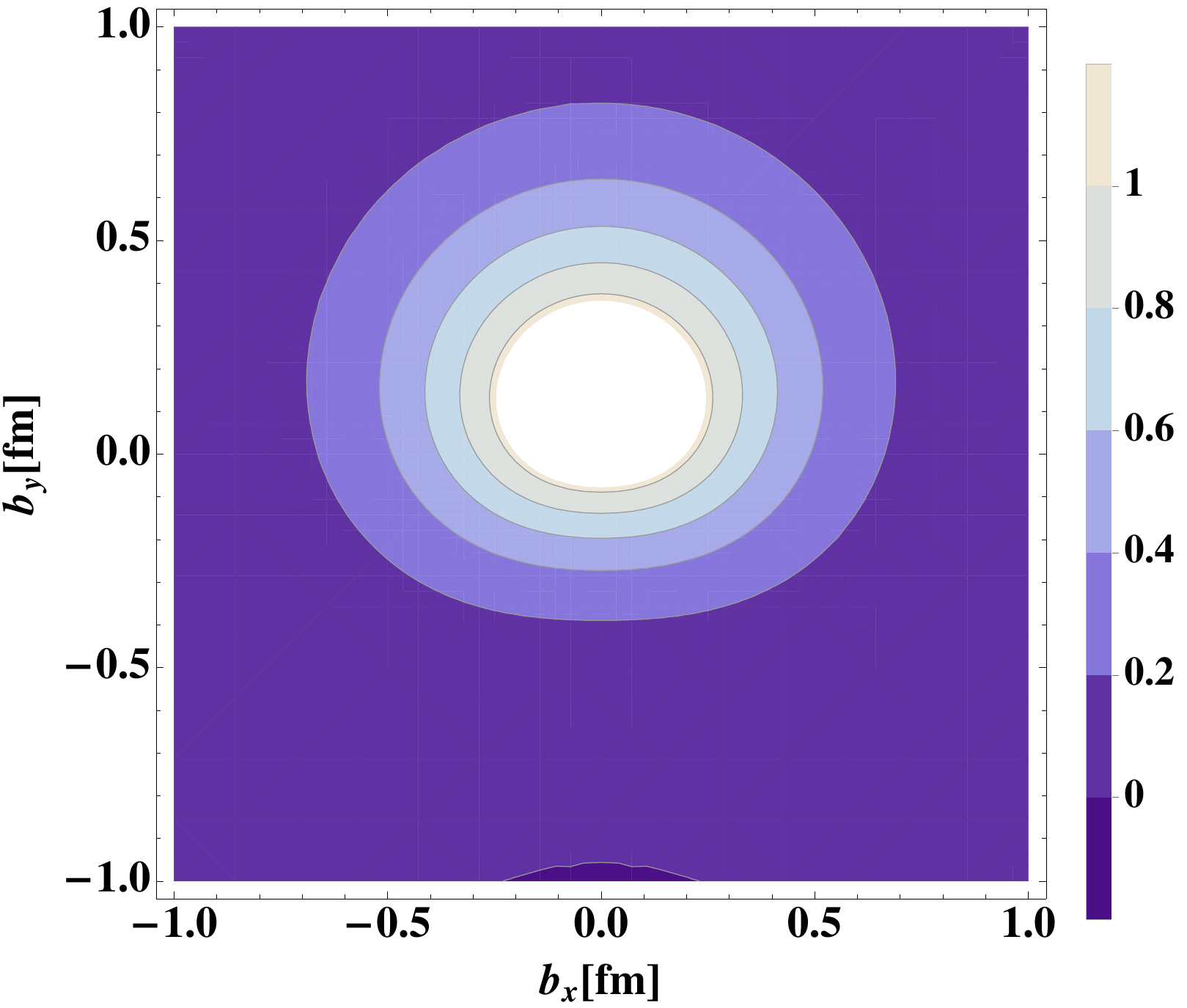}
\end{minipage}
\caption{\label{bcu}Plots of charge densities in the transverse plane for  (a) unpolarized up quark, (b) transversely polarized up quark,  (c) unpolarized down quark, and (d) transversely polarized down quark. The transverse polarization is taken along the $x$ direction.}
\end{figure}

We investigated the quark transverse charge densities inside the unpolarized and transversely polarized nucleon  in Figs. \ref{cmp}(c)$-$\ref{cmp}(d), since they reveal information about the  inner structure of the nucleon. The up quark transverse charge density inside the transversely polarized nucleon is shown to be shifted to the negative $b_y$ direction, while  for the down quark it is distorted in the positive $b_y$ direction.  The down quark is found to be more  influenced due to the transverse polarization of the nucleon.  We have also plotted the top panel of the charge densities in the transverse plane for the unpolarized and  polarized proton and neutron  in Figs. \ref{bcp}(a)$-$\ref{bcp}(d) for the sake of completeness. Also, we  presented  the top view for the up and down quark charge densities in the transverse plane for both the unpolarized and transversely polarized nucleon  in Figs. \ref{bcu}(a)$-$\ref{bcu}(d).
%view from the above

%%%%%%%%%%%%%%%%%%%%%%%%%%%%%%%%%%%%%%%%%%%%%%%%%%%%%%%%%

\section{Summary and conclusion}
\label{sum}
In this work, we have presented a numerical analysis of the helicity independent generalized parton distributions (GPDs) for the  nucleon in the known formalism of the soft-wall model with the inclusion of high Fock states. This approach is based upon the  light-front holography principle to match the matrix elements for nucleon electromagnetic  form factors in AdS modes with the sum rules in QCD that relate the GPDs with form factors. We have presented  the explicit results for the up and down quark GPDs and their Mellin moments in the momentum space. We investigated the GPDs in the impact parameter space as the Fourier transform of GPDs to transverse position space gives access to the distribution of partons in the transverse plane.   It has been observed that the magnitude of GPDs $H(x,Q^2)$ and $q(x,b)$  are larger for the up quark than down quark, whereas for GPDs  $E(x,Q^2)$ and $e^q(x,b)$ the magnitudes  are comparable for up and down quarks. 

We  calculated the charge and anomalous magnetization densities for the unpolarized and transversely polarized nucleons as the transverse charge densities give us important information about the spatial distribution of partons in the transverse plane. The unpolarized nucleon densities are symmetric in the transverse plane, whereas  they become distorted for  the transversely polarized nucleon.  In the case of a proton polarized  transversely along the positive $x$ direction, the corresponding transverse charge density is shifted to the negative $b_y$ direction. In the case of neutron polarized transversely along the $x$ axis, the positive charges move towards the positive $b_y$ direction while the negative charges are forced to the negative $b_y$ direction.  We have also performed the flavor decomposition of the transverse charge densities inside the polarized nucleon for up and down quarks.  The up quark transverse charge density for the nucleon transversely polarized along the negative $x$ axis is found to be shifted to the positive $b_y$ direction while the down quark is more distorted in the opposite directiona; however, the distortion in the down quark is found to be much stronger  than in the up quark.  The overall agreement between the AdS/QCD predictions and the Kelly parametrization approach is remarkable.

\section{Acknowledgements}
N.S. would like to thank S.J. Brodsky for initiating this work and for useful discussions. This work is supported by Department of Science and Technology, Government of India (Grant No. SR/FTP/PS-057/2012).

\end{document}